\documentclass[showkeys,amsmath,amssymb,aps,groupedaddress,floatfix]{revtex4}

\pdfoutput=1

\usepackage{graphicx}
\graphicspath{{figs/}}
\usepackage[colorlinks,bookmarksopen,bookmarksnumbered,citecolor=black,urlcolor=black,linkcolor=black]{hyperref}
\usepackage[utf8]{inputenc}
\usepackage{fancyvrb}
\usepackage{verbatim}
\usepackage{xspace}

\makeatletter
\newcommand{\verbatimfont}[1]{\def\verbatim@font{#1}}%
\makeatother

\verbatimfont{\small\sf\ttfamily}

\newcommand{\vc}[1]{\ensuremath{\vec{\textbf{#1}}}}
\newcommand{\ofrt}{\ensuremath{\left(\vc{r},t \right)}}
\newcommand{\m}{\vc{m}\xspace}   
\newcommand{\M}{\vc{M}\xspace}   
\newcommand{\Msat}{\ensuremath{M_\mathrm{sat}}}
\newcommand{\Bsat}{\ensuremath{B_\mathrm{sat}}}
\newcommand{\B}[1]{\vc{B}\ensuremath{_\mathrm{#1}}}   
\newcommand{\Beff}{\B{eff}}   
\newcommand{\tq}[1]{\vc{\ensuremath{\tau}}\ensuremath{_\mathrm{#1}}}   
\newcommand{\damp}{\ensuremath{\alpha}\xspace}
\newcommand{\Kern}{\hat{{\textbf{K}}}}

\newcommand{\hspin}{(\vc{u}\cdot\nabla)\vc{m}} 
\newcommand{\diff}[2]{\frac{\partial #1}{\partial #2}}
\newcommand{\um}{\,$\mu$m\xspace}
\newcommand{\nm}{\,nm\xspace}
\newcommand{\x}{\,$\times$\,}
\newcommand{\Aex}{$A_\mathrm{ex}$}
\newcommand{\Dex}{$D_\mathrm{ex}$}
\newcommand{\Ku}{$K_\mathrm{u1}$}

\newcommand{\E}[1]{$\times$10$^{#1}$}
\newcommand{\Edens}[1]{\ensuremath{\mathcal{E}_\mathrm{#1}}}
\newcommand{\Fig}[1]{Fig.\,\ref{#1}}

\newcommand{\code}[1]{\texttt{#1}}
\newcommand{\etal}{\textit{et al.}\xspace}
\newcommand{\mumax}{\textsc{MuMax$^3$}\xspace}
\newcommand{\mumag}{$\mu$Mag\xspace}
\newcommand{\UGent}{DyNaMat LAB, Department of Solid State Sciences, Ghent University, Belgium}

\begin{document}
\setlength{\parindent}{0em}

\author{Arne Vansteenkiste$^1$, 
        Jonathan Leliaert$^1$, 
        Mykola Dvornik$^1$, 
        Felipe Garcia-Sanchez$^{2,3}$ 
        and Bartel Van Waeyenberge$^1$\\\vspace{0.1cm}
        \small{$^1$\UGent\\ $^2$Institut d’Electronique Fondamentale Univ. Paris-Sud, 91405 Orsay, France\\ $^3$UMR 8622, CNRS, 91405 Orsay, France}}\normalsize

\email{Arne.Vansteenkiste@UGent.be}

\title{The design and verification of Mumax3}

\begin{abstract}
We report on the design, verification and performance of \mumax, an open-source GPU-accelerated micromagnetic simulation program. This software solves the time- and space dependent magnetization evolution in nano- to micro scale magnets using a finite-difference discretization. Its high performance and low memory requirements allow for large-scale simulations to be performed in limited time and on inexpensive hardware. We verified each part of the software by comparing results to analytical values where available and to micromagnetic standard problems. \mumax also offers specific extensions like MFM image generation, moving simulation window, edge charge removal and material grains.
\end{abstract}

\maketitle

%\tableofcontents

\section{Introduction}

\mumax is a GPU-accelerated micromagnetic simulation program. It calculates the space- and time-dependent magnetization dynamics in nano- to micro-sized ferromagnets using a finite-difference discretization. A similar technique is used by the open-source programs OOMMF\cite{oommf} (CPU) and MicroMagnum\cite{micromagnum} (GPU), and the commercial GpMagnet\cite{Lopez-Diaz2012} (GPU).\\

\mumax is open-source software written in Go\cite{go} and CUDA\cite{cuda}, and is freely available under the GPLv3 license on \url{http://mumax.github.io}. In addition to the terms of the GPL, we kindly request that any work using \mumax refers to the latter website and this paper. An nVIDIA GPU and a Linux, Windows or Mac platform is required to run the software. Apart from nVIDIA's GPU driver, no other dependencies are required to run \mumax.\\

Finite-element micromagnetic software exists as well like, e.g., NMag\cite{nmag}, TetraMag\cite{Kakay2010}, MagPar\cite{scholz03} and FastMag\cite{Chang2011}.  They offer more geometrical flexibility than finite-difference methods, at the expense of performance.\\

%Our main design goal is simplicity and correctness. At present \mumax consists of only about 15\,000 lines of code. Most of it is written in Go, a highly safe programming language. There are only about 1\,500 lines of C for CUDA, which are the only real error-prone parts. Also, a automated tests check over 300 \mumax results against known values on a daily basis.
%Secondly \mumax is designed for low memory usage, motivated by the relatively low amounts GPU memory currently available. 

In this paper we first describe each of \mumax's components and assert their individual correctness and accuracy. Then we address the micromagnetic standard problems \cite{mumag}, where all software components have to work correctly together to solve real-world simulations. We typically compare against OOMMF\cite{oommf} which has been widely used and profoundly tested for over more than a decade. Finally, we report on the performance in terms of speed and memory consumption.\\

The complete input files used to generate the graphs in this paper are available in appendix \ref{appendixA}, allowing for each of the presented results to be reproduced independently. The scripts were executed with \textsc{MuMax} version 3.6.\\

\clearpage

\section{Design}

\subsection{Material Regions}

\begin{figure}
\includegraphics[width=0.5\linewidth]{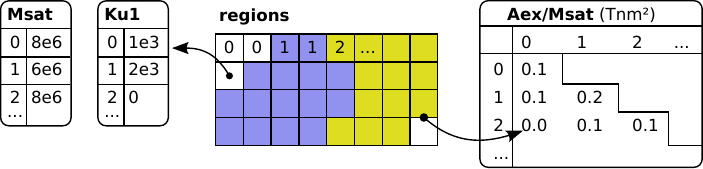}
\caption{Each simulation cell is attributed a region index representing the cell's material type. Material parameters like the saturation magnetization \Msat, anisotropy constants, etc are stored in 1D look-up tables indexed by the region index. Interfacial parameters like the exchange coupling \Aex/\Msat are stored in a 2D lower triangular matrix indexed by the interface's two neighbor region indices.}\label{figRegions}
\end{figure}

\mumax employs a finite difference (FD) discretization of space using a 2D or 3D grid of orthorhombic cells. Volumetric quantities, like the magnetization and effective field, are treated at the center of each cell. On the other hand, interfacial quantities, like the exchange coupling, are considered on the faces in between the cells (\Fig{figRegions}).\\

In order to preserve memory, space-dependent material parameters are not explicitly stored per-cell. Instead, each cell is attributed a \emph{region index} between 0 and 256. Different region indices represent different materials. The actual material parameters are stored in 256-element look-up tables, indexed by the cell's region index.\\

Interfacial material parameters like the exchange coupling are stored in a triangular matrix, indexed by the region numbers of the interacting cells. This allows arbitrary exchange coupling between all pairs of materials (Section \ref{Bexch}). \\

\paragraph*{Time-dependent parameters} In addition to region-wise space-dependence, material parameters in each region can be time-dependent, given by one arbitrary function of time per region.\\

Excitations like the externally applied field or electrical current density can be set region- and time-wise in the same way as material parameters. Additionally they can have an arbitrary number of extra terms of the form $f(t)\times g(x,y,z)$, where $f(t)$ is any function of time multiplied by a continuously varying spatial profile $g(x,y,z)$. This allows to model smooth time- and space dependent excitations like, e.g., an antenna's RF field or an AC electrical current.

\subsection{Geometry}

\mumax uses \emph{Constructive Solid Geometry} to define the shape of the magnet and the material regions inside it. Any shape is represented by a function $f(x,y,z)$ that returns true when $(x,y,z)$ lies inside the shape or false otherwise. E.g. a sphere is represented by the function $x^2+y^2+z^2\leq r^2$. Shapes can be rotated, translated, scaled and combined together with boolean operations like AND, OR, XOR. This allows for complex, parametrized geometries to be defined programmatically. E.g., \Fig{figCSG} shows the magnetization in the logical OR of an ellipsoid and cuboid.

\begin{samepage}
\begin{figure}
\includegraphics[width=0.5\linewidth]{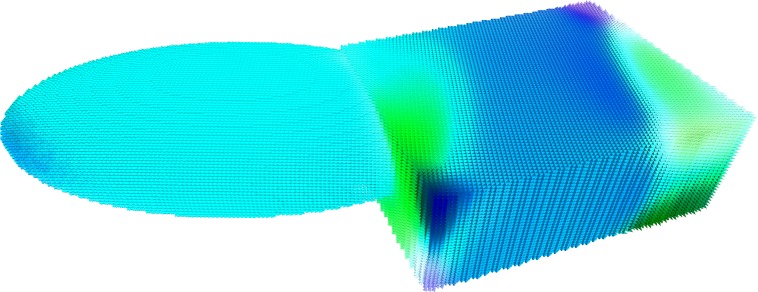}
\caption{Geometry obtained by logically combining an ellipsoid and rotated cuboid. The arrows depict the magnetization direction in this complex shape.}\label{figCSG}
\end{figure}
\end{samepage}

\subsection{Interface}

\paragraph*{Input scripts} \mumax provides a dedicated scripting language that resembles a subset of the Go programming language. The script provides a simple means to define fairly complex simulations. This is illustrated by the code snippet below where we excite a Permalloy ellipse with a 1\,GHz RF field:

\begin{Verbatim}[frame=single]
setgridsize(128, 32, 1)
setcellsize(5e-9, 5e-9, 8e-9)
setGeom(ellipse(500e-9, 160e-9))

Msat = 860e3
Aex  = 13e-12
alpha= 0.05

m=uniform(1, 0, 0)
relax()

f := 1e9  // 1GHz
A := 0.01 // 10mT
B_ext = vector(0.1, A*sin(2*pi*f*t), 0)
run(10e-9)
\end{Verbatim}

\paragraph*{Programming} The \mumax libraries can also be called from native Go. In this way, the full Go language and libraries can be leveraged for more powerful input generation and output processing than the built-in scripting.\\

\paragraph*{Web interface} \mumax provides web-based HTML\,5 user interface. It allows to inspect and control simulations from within a web browser, whether they are running locally or remotely. Simulations may also be entirely constructed and run from within the web GUI. In any case an input file corresponding to the user's clicks is generated, which may later be used to repeat the simulation in an automated fashion.\\

\paragraph*{Data format} \mumax uses OOMMF's "OVF" data format for input and output of all space-dependent quantities. This allows to leverage existing tools. Additionally a tool is provided to convert the output to several other data formats like paraview's VTK\cite{paraview}, gnuplot\cite{gnuplot}, comma-separated values (CSV), Python-compatible JSON, \ldots, and to image formats like PNG, JPG and GIF. Finally, the output is compatible with the 3D rendering software \textsc{MuView}, contributed by Graham Rowlands\cite{muview}.\\

\section{Dynamical terms}

\mumax calculates the evolution of the reduced magnetization $\m\ofrt$, which has unit length. In what follows the dependence on time and space will not be explicitly written down. We refer to the time derivative of \m as the {torque} \tq{} (units 1/s):

\begin{equation}
	\diff \m t = \tq{} \label{eqDyn}
\end{equation}

\tq{} has three contributions: 
\begin{itemize}
\item Landau-Lifshitz torque \tq{LL} (Section \ref{tqLL})
\item Zhang-Li spin-transfer torque \tq{ZL} (Section \ref{tqZL})
\item Slonczewski spin-transfer torque \tq{SL} (Section \ref{tqSL}).
\end{itemize}

\subsection{Landau-Lifshitz torque}\label{tqLL}

\mumax uses the following explicit form for the Landau-Lifshitz torque \cite{landau35, gilbert55}:

\begin{equation}
	\tq{LL} = \gamma_\mathrm{LL} \frac{1}{1+\damp^2} \left(  \m \times \Beff  +\damp\left( \m \times \left( \m \times \Beff \right)\right)   \right) \label{eqLLG}
\end{equation}

with $\gamma_\mathrm{LL}$ the {gyromagnetic ratio} (rad/Ts), \damp the dimensionless {damping parameter} and \Beff\ the {effective field} (T). The default value for $\gamma_\mathrm{LL}$ can be overridden by the user. \Beff\ has the following contributions:   
\begin{itemize}
\item externally applied field \B{ext}
\item magnetostatic field \B{demag} (\ref{Bdemag})
\item Heisenberg exchange field \B{exch} (\ref{Bexch})
\item Dzyaloshinskii-Moriya exchange field \B{dm} (\ref{Bdm})
\item magneto-crystalline anisotropy field \B{anis} (\ref{Banis})
\item thermal field \B{therm} (\ref{Btherm}).
\end{itemize}

\begin{figure}
\includegraphics[width=0.5\linewidth]{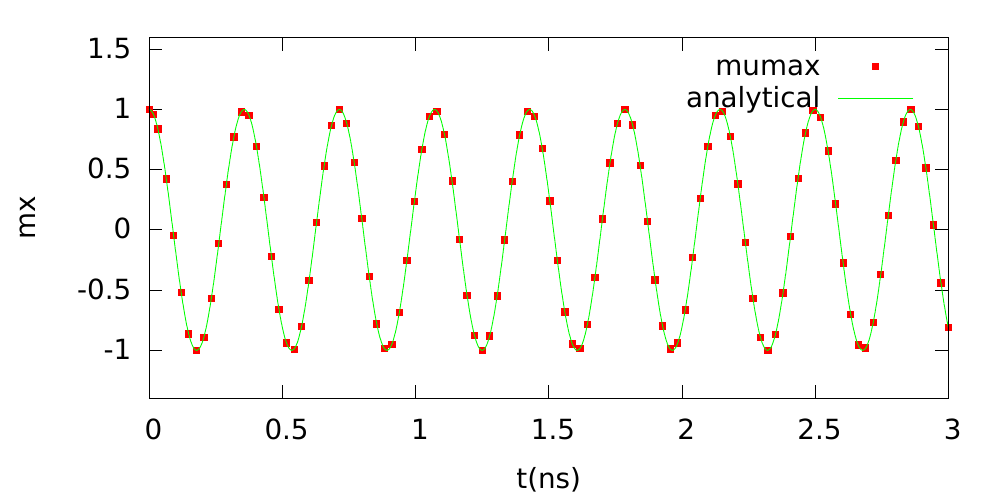}
\caption{Validation of Eq. \ref{eqLLG} for a single spin precessing without damping in a 0.1\,T field along $z$, perpendicular to \vc m. Analytical solution: $m_x = \cos(0.1\mathrm{T} \gamma_{LL} t)$.}\label{figtqLL}
\end{figure}

\Fig{figtqLL} shows a validation of the Landau-Lifshitz torque for a single spin precessing without damping in a constant external field.\\

%{precession.txt}

\subsection{Magnetostatic field}\label{Bdemag}

\paragraph*{Magnetostatic convolution}

A finite difference discretization allows the magnetostatic field to be evaluated as a (discrete) convolution of the magnetization with a demagnetizing kernel $\Kern$:

\begin{equation}
	\B{demag}\ _i = \Kern_{ij} * \M_{j}
\end{equation}

where \M = \Msat \m is the unnormalized magnetization, with \Msat\ the saturation magnetization (A/m). This calculation is FFT-accelerated based on the well-known convolution theorem.  The corresponding energy density is provided as:\\

\begin{equation}
\Edens{demag} = -\frac{1}{2}\vc M \cdot \B{demag}
\end{equation}

\paragraph*{Magnetostatic kernel}
We construct the demagnetizing kernel $\Kern$ assuming constant magnetization\cite{McMichael1999a} in each finite difference cell and we average the resulting \B{demag} over the cell volumes. The integration is done numerically with the number of integration points automatically chosen based on the distance between source and destination cells and their aspect ratios. The kernel is initialized on CPU in double precision, and only truncated to single before transferring to GPU.\\

The kernel's mirror symmetries and zero elements are exploited to reduce storage and initialization time. This results in a 9\x or 12\x decrease in kernel memory usage for 2D and 3D simulations respectively, and is part of the reason for \mumax's relatively low memory requirements (Section \ref{perf}).\\

\paragraph*{Accuracy} The short-range accuracy of $\Kern$ is tested by calculating the demagnetizing factors of a uniformly magnetized cube, analytically known to be -1/3 in each direction. The cube was discretized in cells with varying aspect ratios along $z$ to stress the numerical integration scheme. The smallest possible number of cells was used to ensure that the short-range part of the field has an important contribution. The results presented in Table \ref{tabCube} are accurate to 3 or 4 digits. Standard Problem \#2 (\ref{std2}) is another test sensitive to the short-range kernel accuracy\cite{Donahue2000}.\\

\begin{table}
	\caption{\label{tabCube} Demagnetizing factors ($N_{ij} = H_i/M_j$) calculated for a cube discretized in the smallest possible number of cells with given aspect ratio along $z$. The results lie close to the analytical value of $1/3$, even for very elongated (aspect$>$1) or flat (aspect$<$1) cells. The off-diagonal elements (not shown) are consistent with zero within the single-precision limit.}
	\begin{tabular}{|l||l|l|l|}
		\hline
		\textbf{aspect} & $N\,_{xx}$ &$N\,_{yy}$ & $N\,_{zz}$\\\hline
		8/1  & -0.333207 & -0.333207 & -0.333176 \\
		4/1  & -0.333149 & -0.333149 & -0.333144 \\
		2/1  & -0.333118 & -0.333118 & -0.333118 \\\hline
		1/1  & -0.333372 & -0.333372 & -0.333372 \\\hline
		%1/2  & -0.333118 & -0.333118 & -0.333118 \\
		1/4  & -0.333146 & -0.333146 & -0.333145 \\
		%1/8  & -0.333193 & -0.333193 & -0.333173 \\
		1/16 & -0.333176 & -0.333176 & -0.333280 \\
		%1/32 & -0.333112 & -0.333113 & -0.333471 \\
		1/64 & -0.333052 & -0.333052 & -0.333639 \\
		\hline
	\end{tabular}
\end{table}

\begin{figure}
\includegraphics[width=0.5\linewidth]{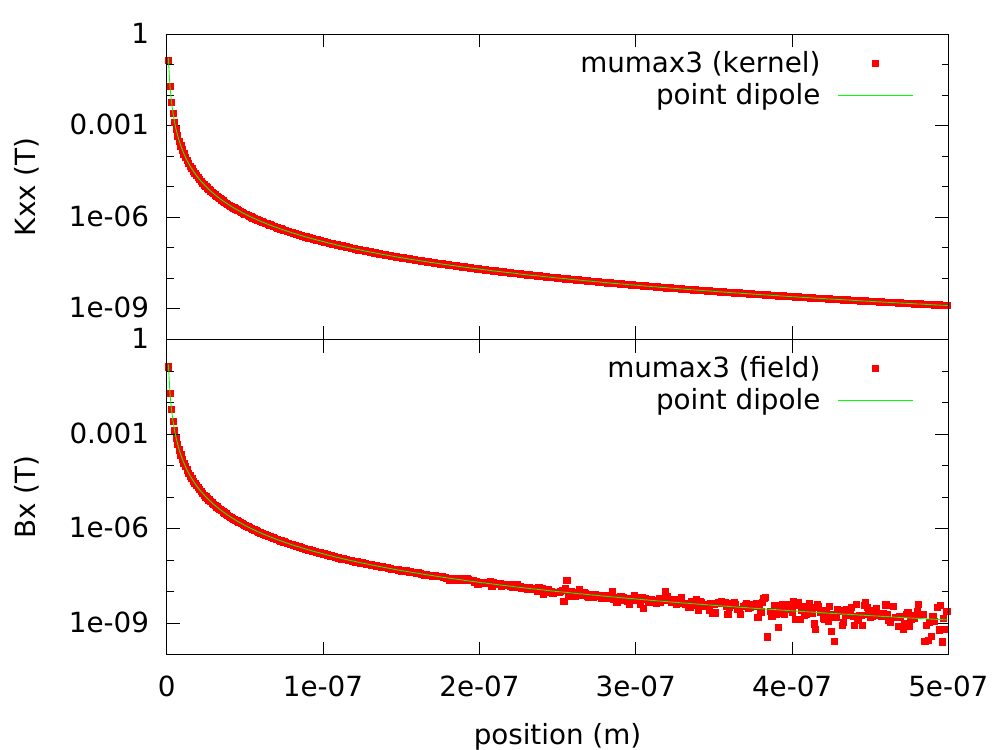}
\caption{\label{figLong} Kernel element $\Kern\ _{xx}$ (top) and $\vec{B}_\mathrm{x}$, the field of a single magnetized cell (1\,nm$^3$, $B_\mathrm{sat}$=1T) (bottom) along the $x$ axis (1\,nm cells), compared to the field of a corresponding dipole. The long-range field remains accurate down to the single-precision numerical limit ($\propto 10^{-7}$\,T).}

\end{figure}

The long-range accuracy of the magnetostatic convolution is assessed by comparing kernel and the field of a single magnetized cell to the corresponding point dipole. The fields presented in \Fig{figLong}, show perfect long-range accuracy for the kernel, indicating accurate numerical integration in that range. The resulting field, obtained by convolution of a single magnetized cell (\Bsat=1\,T) with the kernel, is accurate down to about 0.01\,$\mu$T --- the single-precision noise floor introduced by the FFT's.\\

\paragraph*{Periodic boundary conditions}

\mumax provides optional periodic boundary conditions (PBCs) in each direction. PBCs imply magnetization wrap-around in the periodic directions, felt by stencil operations like the exchange interaction. A less trivial consequence is that the magnetostatic field of repeated magnetization images has to be added to \B{demag}.\\

In contrast to OOMMF's PBC implementation\cite{Lebecki2008}, \mumax employs a so-called macro geometry approach\cite{nmag, Fangohr2009} where a finite (though usually large) number of repetitions is taken into account, and that number can be freely chosen in each direction. \mumax's \code{setPBC(Px, Py, Pz)} command enables $P_x, P_y, P_z$ additional images \emph{on each side} of the simulation box, given that $P$ is sufficiently large.\\

To test the magnetostatic field with PBC's, we calculate the demagnetizing tensors of a wide film and long rod in two different ways: either with a large grid without PBC's, or with a small grid but with PBC's equivalent to the larger grid. In our implementation, a gridsize $(N_x, N_y, N_z)$ with PBC's $(P_x, P_y, P_z)$ should approximately correspond to a gridsize $(2P_xN_x, 2P_yN_y, 2P_zN_z)$ without PBC's. This is verified in tables \ref{tabPBC1} and \ref{tabPBC2} where we extend in plane for the film and along $z$ for the rod. Additionally, for very large sizes both results converge to the well-known analytical values for infinite geometries.\\

\begin{table}
	\caption{\label{tabPBC1} Out-of-plane demagnetizing factors for a thin film with grid size 2\,\x\,2\x\,1 and 2D PBC's \x\,$P$ (column 1) or without PBC's but with a corresponding grid size $\times 2P$ (column 2). Both give comparable results for sufficiently large $P$, verifying the PBC implementation.}
	\begin{tabular}{|l|l||l|l|}\hline
\textbf{PBC}    &  $N_{zz}$  &  \textbf{grid}           &  $N_{zz}$  \\\hline\hline
\x 1      & -0.71257  &  $\times$2      &  -0.76368 \\
%2      & -0.87129  &  $\times$4      &  -0.88483 \\
\x 4      & -0.93879  &  $\times$8      &  -0.94224 \\
%8      & -0.96964  &  $\times$16     &  -0.97051 \\
\x 16     & -0.98438  &  $\times$32     &  -0.98460 \\
%32     & -0.99158  &  $\times$64     &  -0.99163 \\
\x 64     & -0.99514  &  $\times$128    &  -0.99515 \\
%128    & -0.99669  &  $\times$256    &  -0.99691 \\
\x 256    & -0.99713  &  $\times$512    &  -0.99779 \\\hline
%512    & -0.99735  &  $\times$1024   &  -0.99823 \\\hline
$\infty$    & -1        &  $\times\infty$   &  -1       \\\hline
	\end{tabular}
\end{table}

\begin{table}
	\caption{\label{tabPBC2} in-plane demagnetizing factors for a long rod with grid size 1\,\x\,1\x\,2 and 1D PBC's \x\,$P$ (column 1) or without PBC's but with a corresponding grid size $\times 2P$ (column 2). Both give comparable results for sufficiently large $P$, verifying the PBC implementation.}
	\begin{tabular}{|l|l||l|l|}\hline
\textbf{PBC}    &  $N{xx}$ &  \textbf{grid}               &  $N{xx}$    \\\hline\hline
\x 1      & -0.251960   &   $\times$2      &  -0.3331182 \\
%2      & -0.414276   &   $\times$4      &  -0.4367338 \\
\x 4      & -0.476766   &   $\times$8      &  -0.4809398 \\
%8      & -0.494110   &   $\times$16     &  -0.4947033 \\
\x 16     & -0.498280   &   $\times$32     &  -0.4983577 \\
%32     & -0.499275   &   $\times$64     &  -0.4992854 \\
\x 64     & -0.499517   &   $\times$128    &  -0.4995183 \\
%128    & -0.499576   &   $\times$256    &  -0.4995766 \\
\x 256    & -0.499590   &   $\times$512    &  -0.4995911 \\\hline
%512    & -0.499592   &   $\times$1024   &  -0.4995948 \\\hline
$\infty$    & -0.5        &  $\times\infty$   &  -0.5       \\\hline
	\end{tabular}
\end{table}

\subsection{Heisenberg exchange interaction}\label{Bexch}

The effective field due to the Heisenberg exchange interaction \cite{brown63}:

\begin{equation}
	\B{exch} = 2\frac{A_\mathrm{ex}}{\Msat} \Delta \m  \label{eqBexch1}
\end{equation}

is evaluated using a 6-neighbor small-angle approximation\cite{Donahue1998, Donahue2004}:

\begin{equation}
	\B{exch} = 2\frac{A_\mathrm{ex}}{\Msat} \sum_i \frac{(\m_i - \m)}{\Delta_i^2} \label{eqBexch1}
\end{equation}

where $i$ ranges over the six nearest neighbors of the central cell with magnetization \m. $\Delta_i$ is the cell size in the direction of neighbor $i$.\\

At the boundary of the magnet some neighboring magnetizations $\m_i$ are missing. In that case we use the cell's own value \m instead of $\m_i$, which is equivalent to employing Neumann boundary conditions \cite{Donahue1998, Donahue2004}.\\

The corresponding energy density is provided as:\\

\begin{eqnarray}
\Edens{exch} &=& A_\mathrm{ex}(\nabla \vc m)^2\label{eqEdensGrad}\\
&=& -\frac{1}{2}\vc M \cdot \B{exch}\label{eqEdensExch}
\end{eqnarray}

\begin{figure}
\includegraphics[width=0.5\linewidth]{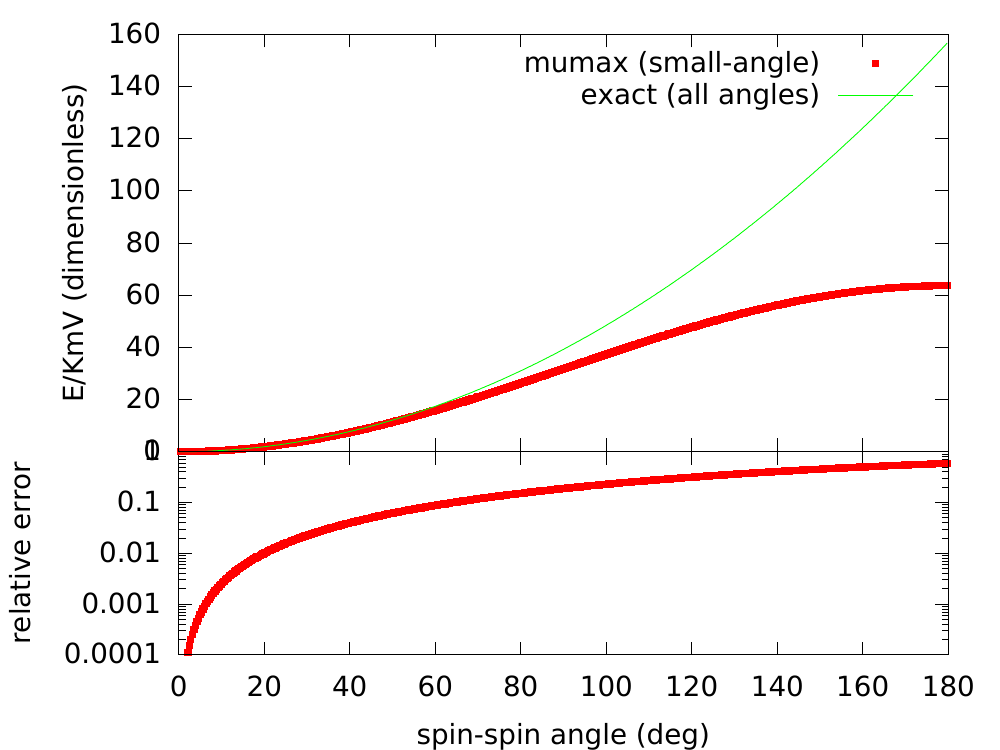}
\caption{Numerical (Eq.\ref{eqBexch1},\ref{eqEdensExch}) and analytical (Eq.\ref{eqEdensGrad}) exchange energy denisty (in units $K_m=1/2\mu_0 M_\mathrm{sat}^2$) for spiral magnetizations as a function of the angle between neighboring spins (independent of material parameters). To ensure an accurate energy, spin-spin angles should be kept below $\propto$20--30$^\circ$ by choosing a sufficiently small cell size.} \label{figExchE}
\end{figure}

\mumax calculates the energy from the effective field using Eqns. \ref{eqBexch1}, \ref{eqEdensExch}. The implementation is verified by calculating the exchange energy of a 1D magnetization spiral, for which the exact form (Eq.\ref{eqEdensGrad}) is easily evaluated. \Fig{figExchE} shows that the linearized approximation is suited as long as the angle between neighboring magnetizations is not too large. This can be achieved by choosing a sufficiently small cell size compared to the exchange length.\\

\paragraph*{Inter-region exchange}

The exchange interaction between different materials deserves special attention. \Aex\ and \Msat\ are defined in the cell volumes, while Eq. \ref{eqBexch1} requires a value of \Aex/\Msat\ properly averaged out between the neighboring cells. For neighboring cells with different material parameters \Aex$_1$, \Aex$_2$ and \Msat$_1$, \Msat$_2$  \mumax uses a harmonic mean:

\newcommand{\lex}[1]{\frac{A_\mathrm{ex#1}}{M_\mathrm{sat#1}}}
\begin{equation}
	\B{exch} = 2S\frac{2\lex{1}\lex{2}}{\lex{1}+\lex{2}} \sum_i \frac{(\m_i - \m)}{\Delta_i^2} \label{eqBexch}
\end{equation}

which can easily be derived, and where we set $S=1$ by default. $S$ is an arbitrary scaling factor which may be used to alter the exchange coupling between regions, e.g., to lower the coupling between grains or antiferromagnetically couple two layers.

\subsection{Dzyaloshinskii-Moriya interaction}\label{Bdm}

\mumax provides induced Dzyaloshinskii-Moriya interaction for thin films with out-of-plane symmetry breaking according to \cite{Bogdanov2001}, yielding an effective field term:

\begin{equation}
\B{DM} = \frac{2D}{\Msat} \left(\frac{\partial m_z}{\partial x},\ \frac{\partial m_z}{\partial y},\ -\frac{\partial m_x}{\partial x}-\frac{\partial m_y}{\partial y}\right)\label{eqBDMI}
\end{equation}

where we apply boundary conditions\cite{Rohart2013}:

\begin{eqnarray}
 \left.\frac{\partial m_z}{\partial x}\right|_{\partial V} &=& \frac{D}{2A}m_x\label{eqDMIBC1} \\
 \left.\frac{\partial m_z}{\partial y}\right|_{\partial V} &=& \frac{D}{2A}m_y\\
 \left.\frac{\partial m_x}{\partial x}\right|_{\partial V} = \left.\frac{\partial m_y}{\partial y}\right|_{\partial V} &=& -\frac{D}{2A}m_z \\
 \left.\frac{\partial m_x}{\partial y}\right|_{\partial V} = \left.\frac{\partial m_y}{\partial x}\right|_{\partial V} &=&  0\\
 \left.\frac{\partial m_x}{\partial z}\right|_{\partial V} = \left.\frac{\partial m_y}{\partial z}\right|_{\partial V} = \left.\frac{\partial m_z}{\partial z}\right|_{\partial V} &=&  0\label{eqDMIBC2}
\end{eqnarray}

Numerically, all derivatives are implemented as central derivatives, i.e., the difference between neighboring magnetizations over their distance in that direction: $\partial\m / \partial i = (\m_{i+1} - \m_{i-1})/(2\Delta_i)$. When a neighbor is missing at the boundary ($\partial V$), its magnetization is replaced by $\m + \frac{\partial m}{\partial i}|_{\partial V} \Delta_i\vc n$ where $\m$ refers to the central cell and the relevant partial derivative is selected from Eq. \ref{eqDMIBC1}--\ref{eqDMIBC2}.\\

In case of nonzero $D$, these boundary conditions are simultaneously applied to the Heisenberg exchange field.\\

The effective field in Eq.\ref{eqBDMI} gives rises to an energy density:

\begin{eqnarray}
\Edens{exch(DM)} &=& m_z(\nabla\cdot\vc m) - (\vc m\cdot\nabla)m_z\label{eqEdDMIex}\\
&=& -\frac{1}{2}\vc M \cdot \B{exch(DM)}\label{eqEdensDMI}
\end{eqnarray}

Similar to the anisotropic exchange case, \mumax calculates the energy density from Eqns.\ref{eqEdensDMI}, \ref{eqBDMI}. Eq.\ref{eqEdDMIex} is the exact form, well approximated for sufficiently small cell sizes.\\

\begin{figure}
\includegraphics[width=0.5\linewidth]{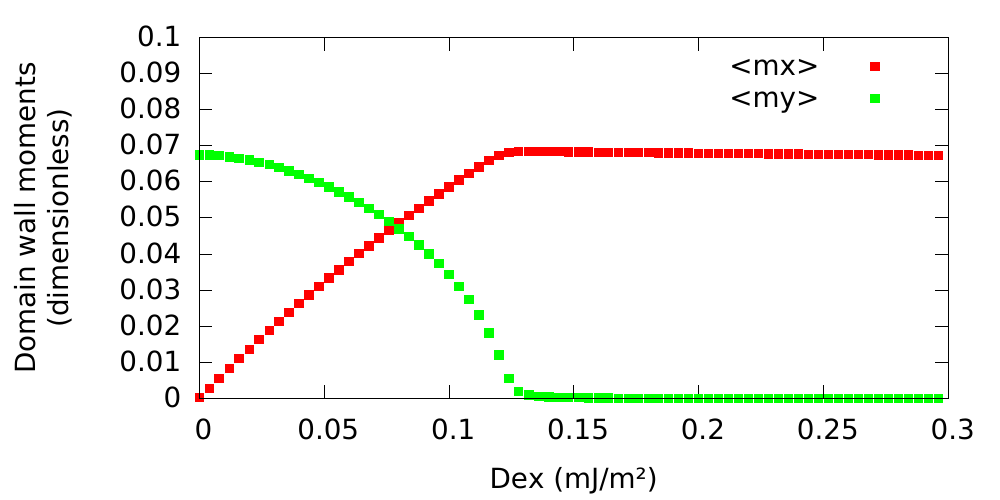}
\caption{\label{figThiaville} Simulated domain wall magnetization in a 250\,nm wide, 0.6\,nm thick Co/Pt film (\Msat=1100\E3A/m, \Aex=16\E{-12}J/m, $K_\mathrm{u1}$=1.27\E6 J/m$^3$) as a function of the Dzyaloshinskii-Moriya strength \Dex. The left-hand and righ-hand sides correspond to a Bloch and Néel wall, respectively. Results correspond well to \cite{Thiaville2012}.}
\end{figure}

In \Fig{figThiaville}, the DMI implementation is compared to the work of Thiaville \etal\cite{Thiaville2012}, where the transformation of a Bloch wall into a Néel wall by varying \Dex\ is studied.

\subsection{Magneto-crystalline anisotropy}\label{Banis}

\paragraph*{Uniaxial}

\mumax provides uniaxial magneto-crystalline anisotropy in the form of an effective field term:

\begin{eqnarray}
\B{anis} &=&   \frac{2 K_\mathrm{u1}}{\Bsat} (\vc u \cdot \vc m)   \vc u\nonumber\\
         &+&   \frac{4 K_\mathrm{u2}}{\Bsat} (\vc u \cdot \vc m)^3 \vc u 
\end{eqnarray}

where $K_\mathrm{u1}$ and $K_\mathrm{u2}$ are the first and second order uniaxial anisotropy constants and $\vc u$ a unit vector indicating the anisotropy direction. This corresponds to an energy density:
\begin{eqnarray}
\mathcal{E}_\mathrm{anis} &=& -K_\mathrm{u1}(\vc u \cdot \vc m)^2- K_\mathrm{u2}(\vc u\cdot \vc m)^4 \label{EanisUA} \\
&=& -\frac{1}{2} \B{anis}(K_\mathrm{u1})\cdot\vc M -\frac{1}{4} \B{anis}(K_\mathrm{u2})\cdot\vc M \label{EaniUM}
\end{eqnarray}

\begin{figure}
\includegraphics[width=0.22\linewidth]{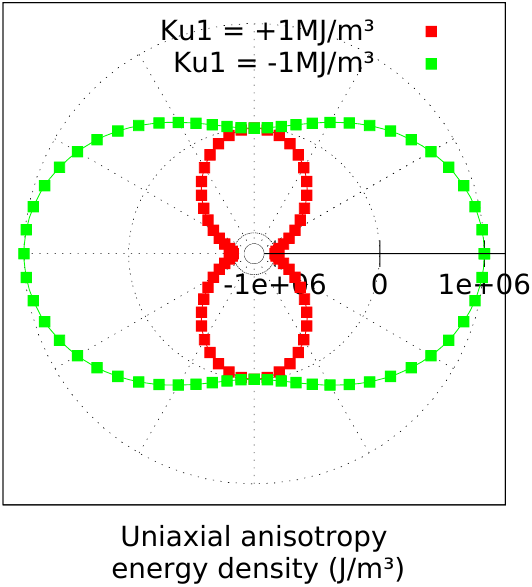}
\includegraphics[width=0.22\linewidth]{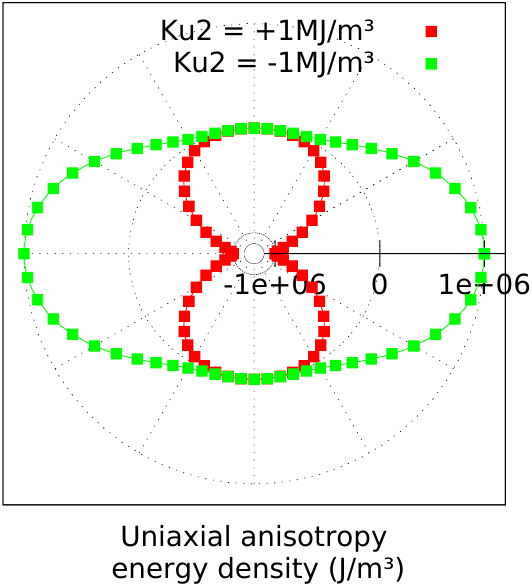}\\
\includegraphics[width=0.15\linewidth]{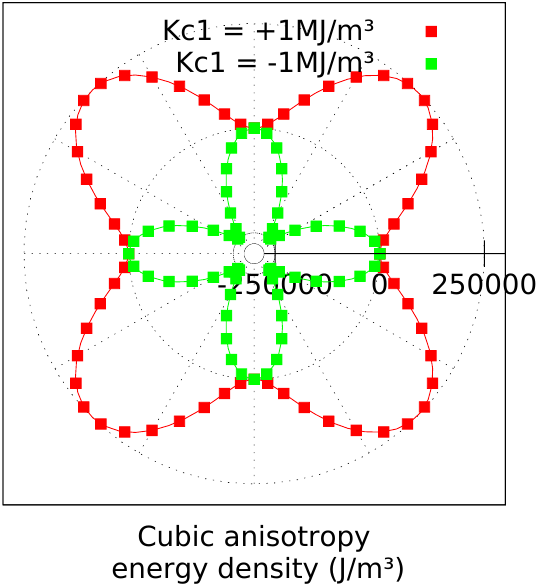}
\includegraphics[width=0.15\linewidth]{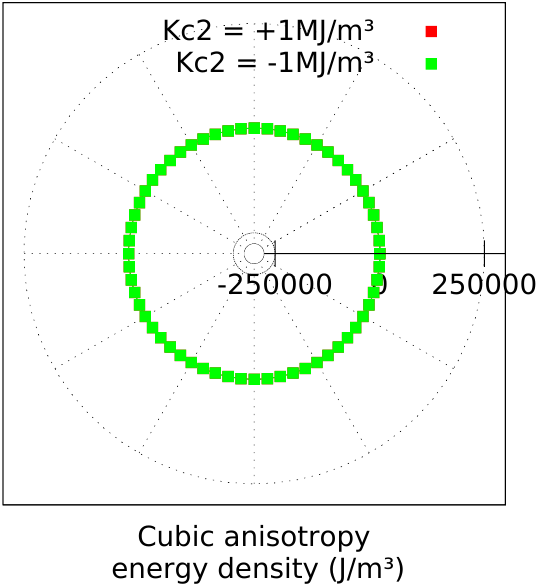}
\includegraphics[width=0.15\linewidth]{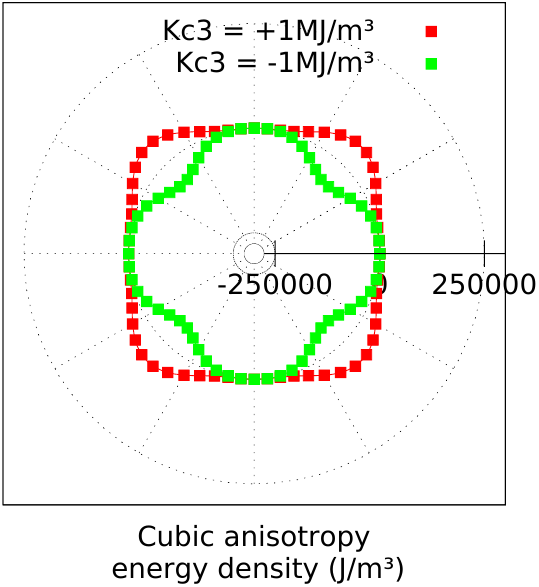}
\caption{\label{figAnis}Uniaxial (top) and cubic (bottom) anisotropy energy density of a single spin as a function of its orientation in the $xy$-plane. The uniaxial axis is along $x$, the cubic axes along $x$, $y$ and $z$. The dots are computed with \mumax (Eq.\ref{EaniUM},\ref{EanisCM}), lines are analytical expressions (Eq.\ref{EanisUA},\ref{EanisCA}). Positive and negative $K$ values denote hard and easy anisotropy, respectively.}
\end{figure}

\mumax calculates the energy density from the effective field using Eq.\ref{EaniUM}, where $\B{anis}(K_\mathrm{ui})$ denotes the effective field term where only $K_\mathrm{ui}$ is taken into account. The resulting energy is verified in \Fig{figAnis}. Since the energy is derived directly form the effective field, this serves as a test for the field as well.\\

\paragraph*{Cubic}

\mumax provides cubic magneto-crystalline anisotropy in the form of an effective field term:

\begin{eqnarray}
\B{anis} =&&\nonumber\\
- 2 K_\mathrm{c1}/\Msat(&((\vc c_2\cdot \vc m)^2 + (\vc c_3\cdot \vc m)^2 ) ( (\vc c_1\cdot \vc m) \vc c_1)&+   \nonumber\\ 
                        &((\vc c_1\cdot \vc m)^2 + (\vc c_3\cdot \vc m)^2 ) ( (\vc c_2\cdot \vc m) \vc c_2)&+   \nonumber\\
                        &((\vc c_1\cdot \vc m)^2 + (\vc c_2\cdot \vc m)^2 ) ( (\vc c_3\cdot \vc m) \vc c_3)&)   \nonumber\\
- 2 K_\mathrm{c2}/\Msat(&((\vc c_2\cdot \vc m)^2   (\vc c_3\cdot \vc m)^2 ) ( (\vc c_1\cdot \vc m)   \vc c_1)&+ \nonumber\\
                        &((\vc c_1\cdot \vc m)^2   (\vc c_3\cdot \vc m)^2 ) ( (\vc c_2\cdot \vc m)   \vc c_2)&+ \nonumber\\
                        &((\vc c_1\cdot \vc m)^2   (\vc c_2\cdot \vc m)^2 ) ( (\vc c_3\cdot \vc m)   \vc c_3)&) \nonumber\\
- 4 K_\mathrm{c3}/\Msat(&((\vc c_2\cdot \vc m)^4 + (\vc c_3\cdot \vc m)^4 ) ( (\vc c_1\cdot \vc m)^3 \vc c_1)&+ \nonumber\\
                        &((\vc c_1\cdot \vc m)^4 + (\vc c_3\cdot \vc m)^4 ) ( (\vc c_2\cdot \vc m)^3 \vc c_2)&+ \nonumber\\
                        &((\vc c_1\cdot \vc m)^4 + (\vc c_2\cdot \vc m)^4 ) ( (\vc c_3\cdot \vc m)^3 \vc c_3)&) \nonumber\\
\end{eqnarray}

where $K_{\mathrm{c}n}$ is the $n$th-order cubic anisotropy constant and $\vc c_1$, $\vc c_2$, $\vc c_3$ a set of mutually perpendicular unit vectors indicating the anisotropy directions. (The user only specifies $\vc c_1$ and $\vc c_2$. We compute $\vc c_3$ automatically as  $\vc c_1 \times \vc c_2$.) This corresponds to an energy density:
\begin{eqnarray}
\mathcal{E}_\mathrm{anis} =&& \nonumber\\
K_\mathrm{c1}  &((\vc c_1\cdot \vc m)^2 (\vc c_2\cdot \vc m)^2 &+ \nonumber\\
               &(\vc c_1\cdot \vc m)^2 (\vc c_3\cdot \vc m)^2 &+ \nonumber\\
               &(\vc c_2\cdot \vc m)^2 (\vc c_3\cdot \vc m)^2)&+ \nonumber\\
K_\mathrm{c2}  &(\vc c_1\cdot \vc m)^2 (\vc c_2\cdot \vc m)^2 (\vc c_3\cdot\vc m)^2 &+ \nonumber\\
K_\mathrm{c3}  &((\vc c_1\cdot \vc m)^4 (\vc c_2\cdot \vc m)^4&+  \nonumber\\
               &(\vc c_1\cdot \vc m)^4 (\vc c_3\cdot \vc m)^4&+  \nonumber\\
               &(\vc c_2\cdot \vc m)^4 (\vc c_3\cdot \vc m)^4&) \label{EanisCA}
\end{eqnarray}

which, just like in the uniaxial case, \mumax computes using the effective field:

\begin{eqnarray}
\mathcal{E}_\mathrm{anis} &=& -\frac{1}{4} \B{anis}(K_\mathrm{c1})\cdot\vc M -\frac{1}{6} \B{anis}(K_\mathrm{c2})\cdot\vc M\nonumber\\
&& -\frac{1}{8} \B{anis}(K_\mathrm{c3})\cdot\vc M \label{EanisCM}
\end{eqnarray}

which is verified in \Fig{figAnis}.\\

\subsection{Thermal fluctuations}\label{Btherm}

\mumax provides finite temperature by means of a fluctuating thermal field \B{therm} according to Brown\cite{Brown1963temp}:

\begin{equation}
	\vc{B}_\mathrm{therm} = \vec\eta(\mathrm{step}) \sqrt{ \frac{2\mu_0\alpha k_\mathrm{B} T}{B_\mathrm{sat}\gamma_\mathrm{LL}\Delta V\Delta t}  }
 \end{equation}

where $\alpha$ is the damping parameter, $k_\mathrm{B}$ the Boltzmann constant, $T$ the temperature, \Bsat\ the saturation magnetization expressed in Tesla, $\gamma_\mathrm{LL}$ the gyromagnetic ratio (1/Ts), $\Delta V$ the cell volume, $\Delta t$ the time step and $\vec\eta(\mathrm{step})$ a random vector from a standard normal distribution whose value is changed after every time step.\\

\begin{figure}
\includegraphics[width=0.5\linewidth]{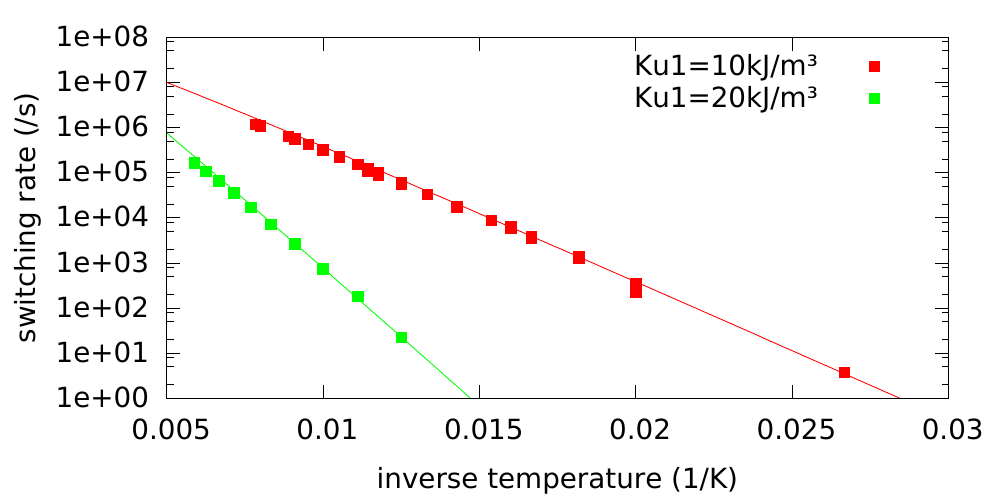}
\caption{\label{figSwitch} Arrhenius plot of the thermal switching rate of a 10\,nm  large cubic particle (macrospin), with \Msat=1MA/m, $\alpha$=0.1, $\Delta t$=\E{-12}s, $K_\mathrm{u1}$=1\E{4} or 2\E{4} J/m$^3$. Simulations were performed on an ensemble of 512$^2$ uncoupled particles for 0.1\,$\mu$s (high temperatures) or 1\,$\mu$s (low temperatures). Solid lines are the analytically expected switching rates (Eq.\ref{eqSwitch}).}
\end{figure}

\paragraph*{Solver constraints} $\vc{B}_\mathrm{therm}$ randomly changes in between time steps. Therefore, only \mumax's Euler and Heun solvers (\ref{solver}) can be used as they do not require torque continuity in between steps. Additionally, with thermal fluctuations enabled we enforce a fixed time step $\Delta t$. This avoids feedback issues with adaptive time step algorithms.\\

\paragraph*{Verification} We test our implementation by calculating the thermal switching rate of a single (macro-)spin particle with easy uniaxial anisotropy. In the limit of a high barrier compared to the thermal energy, the switching rate $f$ is know analytically to be \cite{Breth2012}:

\begin{equation}
f = \gamma_\mathrm{LL}\frac{\alpha}{1+\alpha^2}\sqrt{\frac{8 K_\mathrm{u1}^3 V}{2\pi M_\mathrm{sat}^2kT}}e^{-KV/kT} \label{eqSwitch}
\end{equation}

\Fig{figSwitch} shows Arrhenius plots for the temperature-dependent switching rate of a particle with volume $V$=(10\,nm)$^3$ and $K_\mathrm{u1}$=1\E{4} or 2\E{4} J/m$^3$. The \mumax simulations correspond well to Eq.\ref{eqSwitch}.

\subsection{Zhang-Li Spin-transfer torque} \label{tqZL}

\mumax includes a spin-transfer torque term according to Zhang and Li \cite{Zhang2004}, applicable when electrical current flows through more than one layer of cells:

\begin{eqnarray}
	\nonumber \tq{ZL} &=& \frac{1}{1+\alpha^2} ( \left(1+\xi\alpha\right) \m \times \left(\m \times \hspin \right) +\\
	& & \left(\xi-\alpha\right)\vc{m}\times \hspin ) \\
	\vc{u} &=& \frac{\mu_B \mu_0}{2 e \gamma_0 B_\mathrm{sat}  (1 + \xi^2)} \vc{j}
\end{eqnarray}

where $\vec{j}$ is the current density, $\xi$ is the degree of non-adiabaticity, $\mu_B$ the Bohr magneton and \Bsat\ the saturation magnetization expressed in Tesla.

%Activated whenever j!=0 TODO: don't activate in 2D with only jz != 0.

The validity of our implementation is tested by Standard Problem \#5 (Section \ref{std5}).

\subsection{Slonczewski Spin-transfer torque} \label{tqSL}

\mumax provides a spin momentum torque term according to Slonczewski\cite{Slonczewski1996, Xiao2004}, transformed to the Landau-Lifshitz formalism:

\begin{eqnarray}
	\tq{SL} &=& \beta\frac{\epsilon-\alpha\epsilon'}{1+\alpha^2} (\m \times (\m_P \times \m)) \nonumber\\
				&&- \beta\frac{\epsilon'-\alpha\epsilon}{1+\alpha^2} \m\times \m_P \label{eqSTT}\\
	\beta &=& \frac{j_z \hbar}{ \Msat e d} \\
	\epsilon &=& \frac{P\ofrt \Lambda^2}{(\Lambda^2 + 1)+ (\Lambda^2-1)(\m\cdot\m_P)}
\end{eqnarray}

where $j_z$ is the current density along the $z$ axis, $d$ is the free layer thickness, $\m_P$ the fixed-layer magnetization, $P$ the spin polarization, the Slonczewski $\Lambda$ parameter characterizes the spacer layer, and $\epsilon'$ is the secondary spin-torque parameter.\\

\mumax only explicitly models the free layer magnetization. The fixed layer is handled in the same way as material parameters and is always considered to be on top of the free layer. The fixed layer's stray field is not automatically taken into account, but can be pre-computed by the user and added as a space-dependent external field term.\\

\begin{figure}
\includegraphics[width=0.5\linewidth]{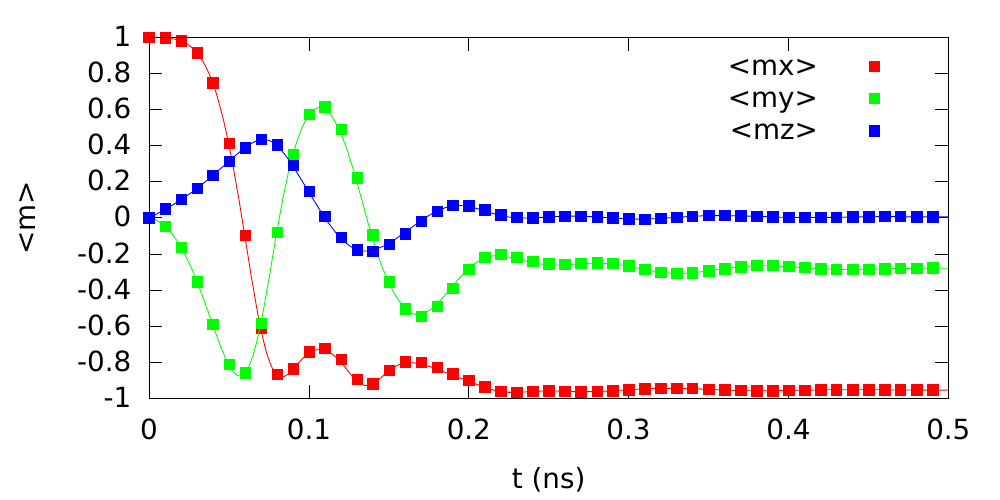}
\caption{\label{figStd5b} Verification of the Slonczewski torque: average magnetization in a 160\,nm\x 80\,nm\x 5\,nm rectangle with \Msat=800\E3\,A/m, \Aex=13\E{-12}\,J/m$^2$, $\alpha$=0.01, $P$ = 0.5669, $j_z$=4.6875\E{11}A, $\Lambda$=2, $\epsilon'$=1, $\vc{m}_p$=(cos(20$^\circ$), sin(20$^\circ$), 0), initial $m$=(1,0,0). Solid line calculated with OOMMF, points by \mumax.}
\end{figure}

As a verification we consider switching an MRAM bit in 160\,nm\x 80\,nm\x 5\,nm Permalloy (\Msat=800\E3\,A/m, \Aex=13\E{-12}\,J/m$^2$, $\alpha$=0.01, $P$ = 0.5669) by a total current of -6\,mA along the $z$ axis using $\Lambda$=2, $\epsilon'$=1. These parameters were chosen so that none of the terms in Eq.\ref{eqSTT} are zero. The fixed layer is polarized at 20$^\circ$ from the $x$ axis to avoid symmetry problems and the initial magnetization was chosen uniform along $x$. The \mumax and OOMMF results shown in \Fig{figStd5b} correspond well.

\section{Time integration}\label{solver}

\subsection{Dynamics}

\mumax provides a number of explicit Runge-Kutta methods for advancing the Landau-Lifshitz equation (Eq.\ref{eqLLG}):

\begin{itemize}
\item  RK45, the Dormand-Prince method, offers 5-th order convergence and a 4-th order error estimate used for adaptive time step control. This is the default for dynamical simulations.
\item  RK32, the Bogacki-Shampine method, offers 3-th order convergence and a 2nd order error estimate used for adaptive time step control. This method is used when relaxing the magnetization to its ground state in which case it performs better than RK45.
\item RK12, Heun's method, offers 2nd order convergence and a 1st order error estimate. This method is used for finite temperature simulations as it does not require torque continuity in between time steps.
\item RK1, Euler's method is provided for academic purposes.
\end{itemize}

\begin{figure}
\includegraphics[width=0.5\linewidth]{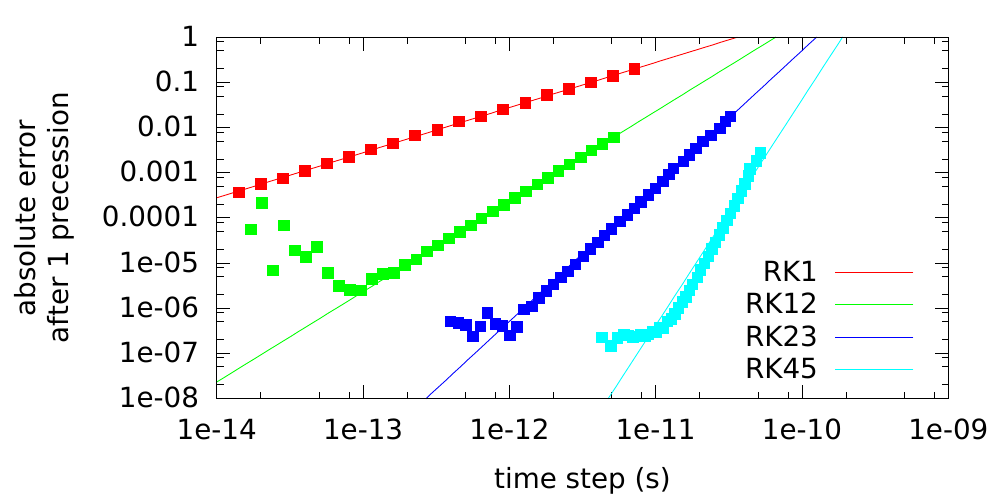}
\caption{\label{figConvergence} Absolute error on a single spin after precessing without damping for one period in a 0.1\,T field, as a function of different solver's time steps. The errors follow 1st, 2nd, 3rd or 5th order convergence (solid lines) for the respective solvers down to a limit set by the single precision arithmetic.}
\end{figure}

These solvers' convergence rates are verified in \Fig{figConvergence}, which serves as a test for their implementation and performance.\\

\paragraph*{Adaptive time step} RK45, RK23 and RK12 provide adaptive time step control, i.e., automatically choosing the time step to keep the error per step $\epsilon$ close to a preset value $\epsilon_0$. As the error per step we use $\epsilon = \mathrm{max}\left|\tau_\mathrm{high}-\tau_\mathrm{low}\right|\Delta t$, with $\tau_\mathrm{high}$ and $\tau_\mathrm{low}$ high-order and low-order torque estimates provided by the particular Runge-Kutta method, and $\Delta t$ the time step. The time step is adjusted using a default headroom of 0.8.

\begin{figure}
\includegraphics[width=0.5\linewidth]{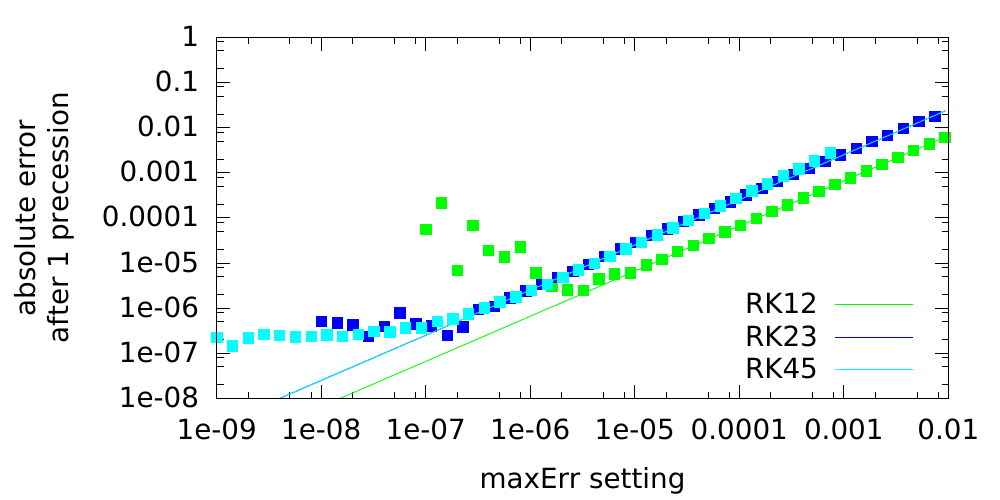}
\caption{\label{figMaxErr} Absolute error on a single spin after precessing without damping for one period in a 0.1\,T field, as a function of different solver's MaxErr settings. Solid lines represent 1st order fits. The same lower bound as in \Fig{figConvergence} is visible.}
\end{figure}

In \mumax, $\epsilon_0$ is accessible as the variable MaxErr. Its default value of 10$^{-5}$ was adequate for the presented standard problems. The relation between $\epsilon_0$ and the overall error at the end of the simulation is in general hard to determine. Nevertheless, we illustrate this in \Fig{figMaxErr} for a single period of spin precession under the same conditions as Fig.\ref{figConvergence}. It can be seen that the absolute error per precession scales linearly with $\epsilon_0$, although the absolute value of the error depends on the solver type and exact simulation conditions.\\%The discrepancy between RK12 and the other methods is attributed to RK12's poor (1st-order) error estimate.

\subsection{Energy minimization}

\mumax provides a \code{relax()} function that attempts to find the systems' energy minimum. This function disables the precession term Eq.\ref{eqLLG}, so that the effective field points towards decreasing energy. \code{Relax} first advances in time until the total energy cuts into the numerical noise floor. At that point the state will be close to equilibrium already. We then begin monitoring the magnitude of the torque instead of the energy, since close to equilibrium the torque will decrease monotonically and is less noisy than the energy. So we advance further until the torque cuts into the noise floor as well. Each time that happens, we decrease \texttt{MaxErr} and continue further until \texttt{MaxErr}=10$^{-9}$. At this point it does not make sense to increase the accuracy anymore (see Fig.\ref{figMaxErr}) and we stop advancing.\\

This \code{Relax} procedure was used in the presented standard problems, where it proved adequate. Typical residual torques after \code{Relax} are of the order of 10$^{-4}$--10$^{-7}$\,$\gamma_\mathrm{LL}$T, indicating that the system is indeed very close to equilibrium. Nevertheless, as with any energy minimization technique, there is always a possibility that the system ends up in a saddle point or very flat part of the energy landscape.\\

\code{Relax} internally uses the RK23 solver, which we noticed performs better then RK45 in most relaxation scenarios. Near equilibrium, both solvers tend to take similarly large time steps, but RK23 needs only half as many torque evaluations per step as RK45.

\section{Standard Problems}

In this section we provide solutions to micromagnetic standard problems \#1--4 provided by the \mumag modeling group\cite{mumag} and standard problem \#5 proposed by Najafi \etal\cite{Najafi2009}. Reference solutions were taken from\cite{mumag} as noted, or otherwise calculated with OOMMF\,1.2\,alpha\,5\,bis\cite{oommf}.

\subsection{Standard Problem \#1}\label{std1}

\begin{figure}
\includegraphics[width=0.5\linewidth]{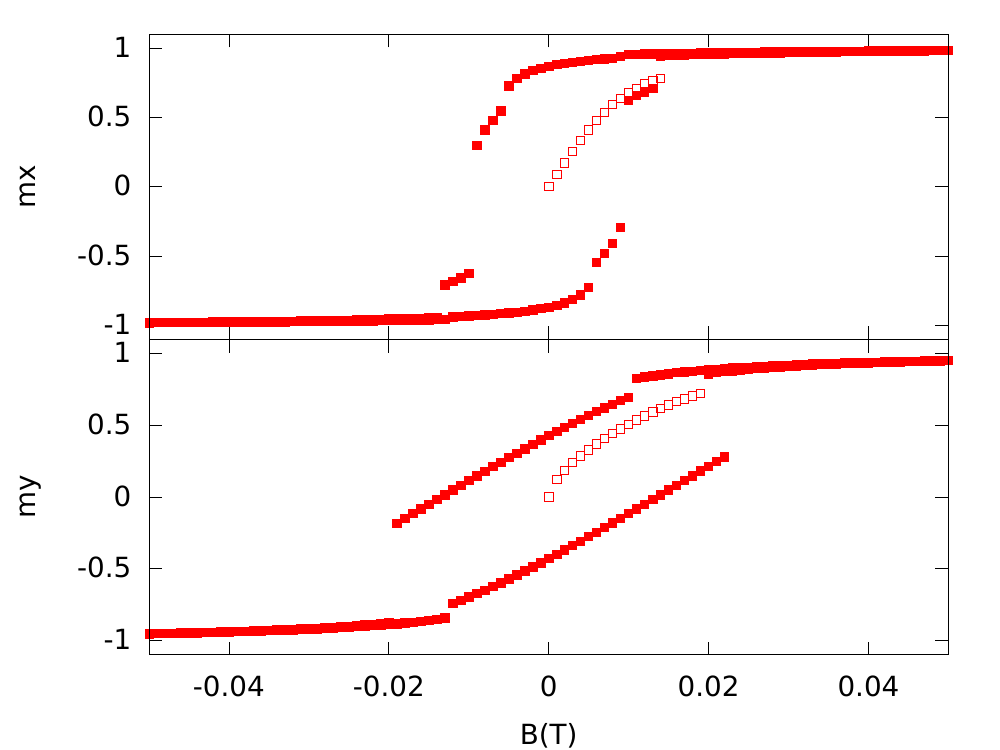}
\caption{\label{figStd1} \mumax solution for standard problem \#1, using a 2D grid of 3.90625\,nm wide cells. Open symbols represent the virgin curve starting from a vortex state. After each field step we applied thermal fluctuations with $\alpha=0.05$, $T=300K$ for 500ps to allow the magnetization to jump over small energy barriers. There are no consistent standard solutions to compare with.}
\end{figure}

The first $\mu$Mag standard problem involves the hysteresis loops of a 1\um\x2\um\x20\nm Permalloy rectangle (\Aex = 1.3\E{-11}J/m, \Msat = 8\E5 A/m, \Ku = 5\E2 J/m$^3$ uniaxial, with easy axis nominally parallel to the long edges of the rectangle) for the field approximately parallel to the long and short axis, respectively. Our solution is presented in Fig.\ref{figStd1}. Unfortunately the submitted solutions\cite{mumag} do not agree with each other, making it impossible to assert the correctness in a quantitative way.

\subsection{Standard Problem \#2}\label{std2}

The second $\mu$Mag standard problem considers a thin film of width $d$, length $5d$ and thickness $0.1d$, all expressed in terms of the exchange length $\l_\mathrm{ex} = \sqrt{2A_\mathrm{ex}/\mu_0M_\mathrm{sat}^2}$. The remanence and coercitive field, expressed in units \Msat, are to be calculated as a function of $d/l_\mathrm{ex}$.\\

\begin{figure}
\includegraphics[width=0.5\linewidth]{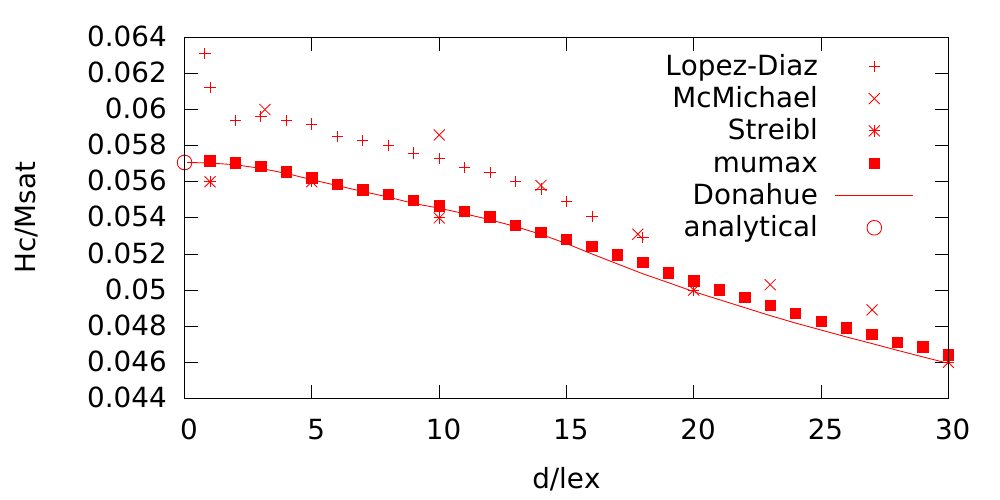}
\caption{\label{figStd2hc}Coercivity for standard problem \#2 as a function of the magnet size $d$ expressed in exchange lengths $l_\mathrm{ex}$. \mumax calculations (points) use automatically chosen cell sizes between 0.25 and 0.5$l_\mathrm{ex}$. OOMMF results (line) taken from\cite{mumag}. The slight discrepancy at high $d$ is attributed to OOMMF's solution using larger cells there. The analytical limit for very small size is by Donahue \etal \cite{Donahue2000}.}
\end{figure}

The coercivity, shown in Fig.\ref{figStd2hc}, behaves interestingly in the small-particle limit where an analytical solution exists\cite{Donahue2000}. In that case the magnetization is uniform and the magnetostatic field dominates the behaviour. Of the solutions submitted to the \mumag group \cite{Streibl1999, Lopez-Diaz1999, McMichael1999,Donahue2000}, the Streibl\cite{Streibl1999}, Donahue\cite{Donahue2000} (OOMMF\,1.1) and \mumax results best approach the small-particle limit. It was shown by Donahue \etal\cite{Donahue2000} that proper averaging of the magnetostatic field over each cell volume is needed to accurately reproduce the analytical limit. Hence this standard problem serves as a test for the numerical integration of our demagnetizing kernel.\\

\begin{figure}
\includegraphics[width=0.5\linewidth]{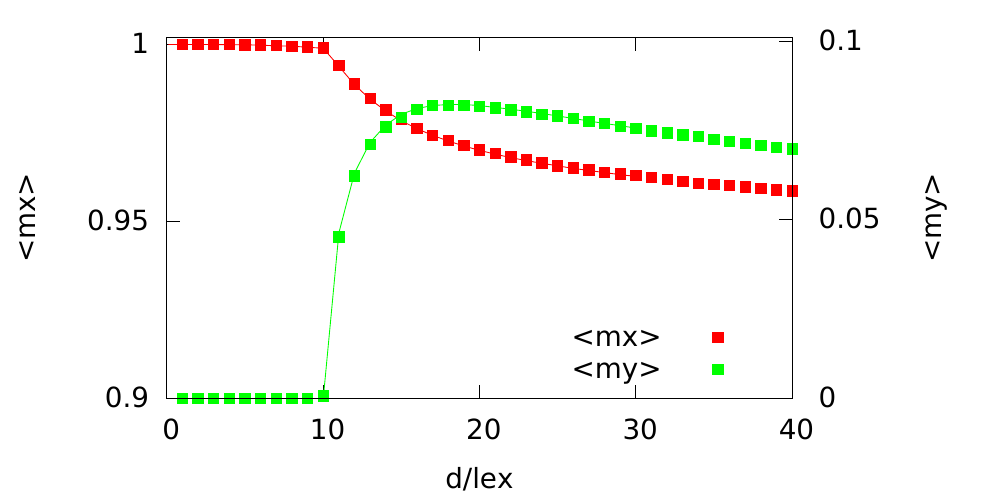}
\caption{\label{figStd2rem} Remanence for standard problem \#2 as a function of the magnet size $d$ expressed in exchange lengths $l_\mathrm{ex}$. The \mumax calculations (points) use automatically chosen cell sizes between 0.25 and 0.5\,$l_\mathrm{ex}$. OOMMF results (line) were taken from\cite{mumag}.}
\end{figure}

\subsection{Standard Problem \#3}\label{std3}

Standard problem \#3 considers a cube with edge length $L$ expressed in exchange lengths $\l_\mathrm{ex} = \sqrt{2A_\mathrm{ex}/\mu_0M_\mathrm{sat}^2}$. The magnet has uniaxial anisotropy with $K_{u1}=0.1 K_m$, with $K_m=1/2\mu_0 M_\mathrm{sat}^2$, easy axis parallel to the $z$-axis. The critical edge size $L$ where the ground state transitions between a quasi-uniform and vortex-like state needs to be found, it is expected around $L$=8.\\

\begin{figure}
\includegraphics[width=0.5\linewidth]{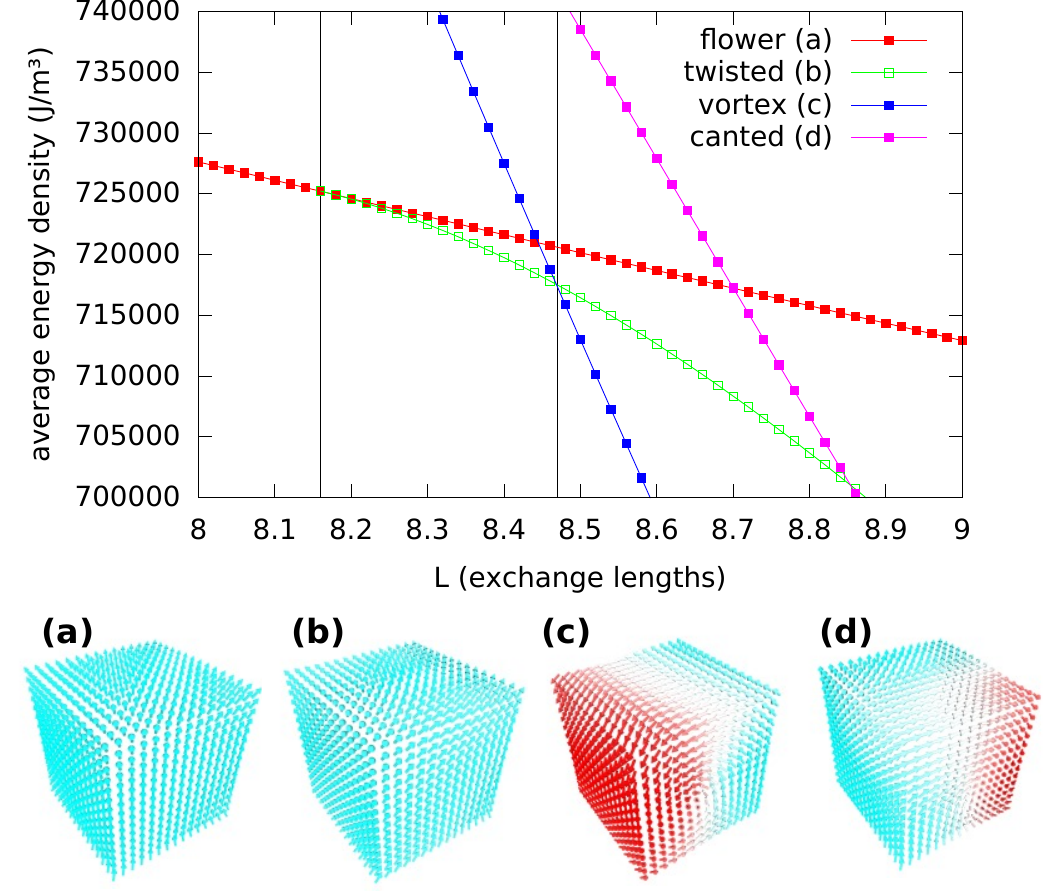}\\
%\tiny{\textbf{(a)}}\includegraphics[width=0.2\linewidth]{std3-flower}
%\tiny{\textbf{(b)}}\includegraphics[width=0.2\linewidth]{std3-twisted}
%\tiny{\textbf{(c)}}\includegraphics[width=0.2\linewidth]{std3-vortex}
%\tiny{\textbf{(d)}}\includegraphics[width=0.2\linewidth]{std3-canted}
\caption{\label{figStd3}Standard problem \#3: energy densities of the flower (a), twisted flower (b), vortex (c) and canted vortex (d) states as a function of the cube edge length $L$. Transitions of the ground state are marked with vertical lines at $L=8.16$ and $L=8.47$.}
\end{figure}

This problem was solved using a 16\x16\x16 grid. The cube was initialized with $\propto$3,000 different random initial magnetization states for random edge lengths $L$ between 7.5 and 9, and relaxed to equilibrium. Four stable states were found, shown in Fig.\ref{figStd3}: a quasi-uniform flower state (a), twisted flower state (b), vortex state (c) and a canted vortex (d). Then cubes of different sizes were initialized to these states and relaxed to equilibrium. The resulting energy for each state, shown in Fig.\ref{figStd3}, reveals the absolute ground states in the considered range: flower state for $L<8.16$, twisted flower for $8.16<L<8.47$ and vortex for $L>8.47$.\\

The transition at $L$=8.47 is in quantitative agreement with the solutions posted to \mumag by Rave \etal \cite{Rave1998} and by Martins \etal \cite{mumag}. The existence of the twisted flower state was already noted by Hertel \etal \cite{mumag}, although without determining the flower to twisted flower transition point.\\

%Fig.\ref{figStd3} shows the phase transition around $L$=8.16, visible as a sudden change in the average $m_z$ as a function of $L$. This is well reproduced by OOMMF simulations (full symbols). The insets showing the spatial magnetization confirm the transition from a quasi-uniform to vortex-like state.\\

%We note that the solutions currently available on the \mumag activity group date from 1998. They report the transition at larger critical edge lengths of $L$=8.47 or $L$=8.52, and report nonzero average $m_y$. Only the solution by Streibl \etal \cite{Streibl1999} provides details on the numerical methods. They computed the magnetostatic field with a finite element method using magnetic charge and asymptotic boundary conditions --- a very different approach than OOMMF and \mumax.  This motivated re-evaluating the problem with a recent OOMMF version and starting the OOMMF simulations from the same state as the \mumax simulations. In that case we find perfect agreement and hence verified the \mumax solution within a finite difference context. A detailed comparison of the accuracy of the FD and FEM approach, or the influence of the exact starting state lies beyond the scope of this work.\\

%This standard problem also specifies the partial energies at the transition point to the vortex state. However, absolute energies are undetermined up to an arbitrary constant. Since \mumax appears to use a different absolute energy offset, this comparison is hard to make.\\

\subsection{Standard Problem \#4}\label{std4}

Standard problem \#4 considers dynamic magnetization reversal in a 500\,nm\x125\,nm\x3\nm Permalloy magnet (\Aex=1.3\E{-11}\,J/m, \Msat=8\E5\,A/m). The initial state is an S-state obtained after saturating along the (1,1,1) direction. Then the magnet is reversed by either field (a): (-24.6, 4.3, 0)\,mT or field (b): (-35.5, -6.3, 0)\,mT. Time-dependent average magnetizations should be given, as well as the space-dependent magnetization when $<m_x>$ first crosses zero.\\

\begin{figure}
\includegraphics[width=0.5\linewidth]{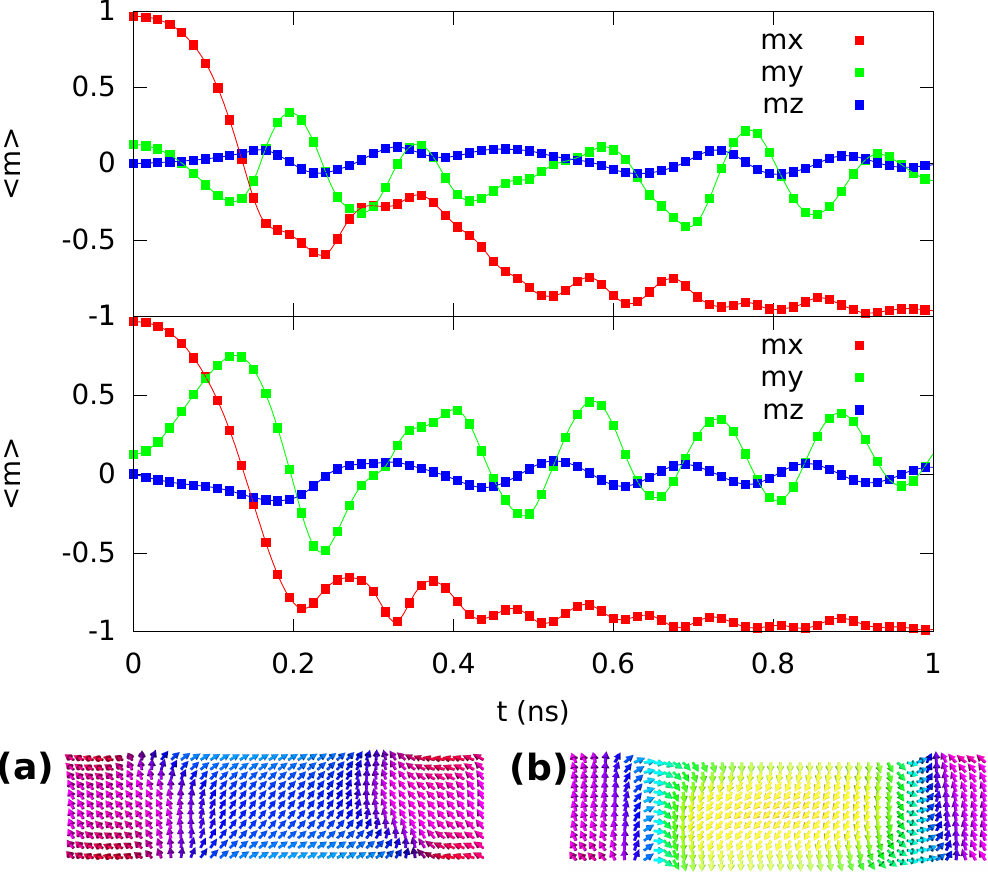}\\
%\tiny{\textbf{(a)}}\includegraphics[width=0.4\linewidth]{std4a}
%\tiny{\textbf{(b)}}\includegraphics[width=0.4\linewidth]{std4b}
\caption{\label{figStd4}\mumax (dots) and OOMMF (lines) solution to standard problem \#4a (top graph) and \#4b (bottom graph), as well as space-dependent magnetization snapshots when $<m_x>$ crosses zero, for fields (a) and (b). All use a 200\x50\x1 grid.}
\end{figure}

Our solution, shown in Fig.\ref{figStd4}, agrees with OOMMF.\\

\subsection{Standard Problem \#5}\label{std5}

\begin{figure}
\includegraphics[width=0.5\linewidth]{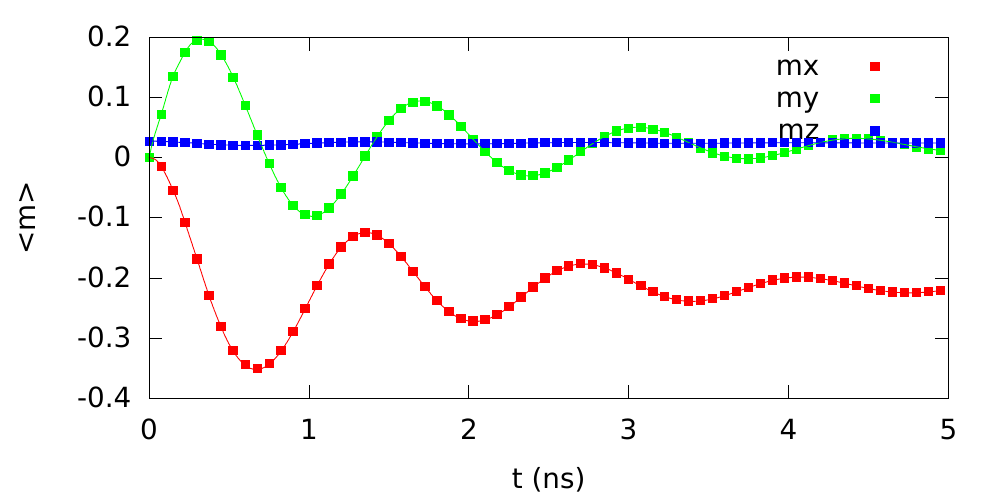}
\caption{\label{figStd5} \mumax (dots) and OOMMF (lines) solution to standard problem \#5, both using a 50\x50\x5 grid.}
\end{figure}

Standard problem \#5 proposed by Najafi \etal\cite{Najafi2009} considers a 100\,nm\,$\times$\,100\,nm\,$\times$\,10\,nm Permalloy square ($A=13\times10^{-12}$\,J/m,  $\Msat=8\times10^5$\,A/m, $\alpha$=0.1, $\xi$=0.05) with an initial vortex magnetization. A homogeneous current $\mathbf{j}=10^{12}$\,Am$^{-2}$ along $x$, applied at $t=0$ drives the vortex towards a new equilibrium state. The obtained time-dependent average magnetization, shown in Fig.\ref{figStd5}, agrees well the OOMMF solution.

\section{Extensions}

\mumax is designed to be modular and extensible. Some of our extensions, described below, have been merged into the mainline code because they may be of general interest. Nevertheless, extensions are considered specific to certain needs and are less generally usable than the aforementioned main features. E.g., MFM images and Voronoi tessellation are only implemented in 2D and only qualitatively tested.\\

\subsection{Moving frame}

\mumax provides an extension to translate the magnetization with respect to the finite difference grid (along the $x$-axis), inserting new values from the side. This allows the simulation window to seemingly "follow" a region of interest like domain wall moving in a long nanowire, without having to simulate the entire wire. \mumax can automatically translate the magnetization to keep an average magnetization component of choice as close to zero as possible, or the user may arbitrarily translate $\vc m$ from the input script.\\

\begin{figure}
\includegraphics[width=0.5\linewidth]{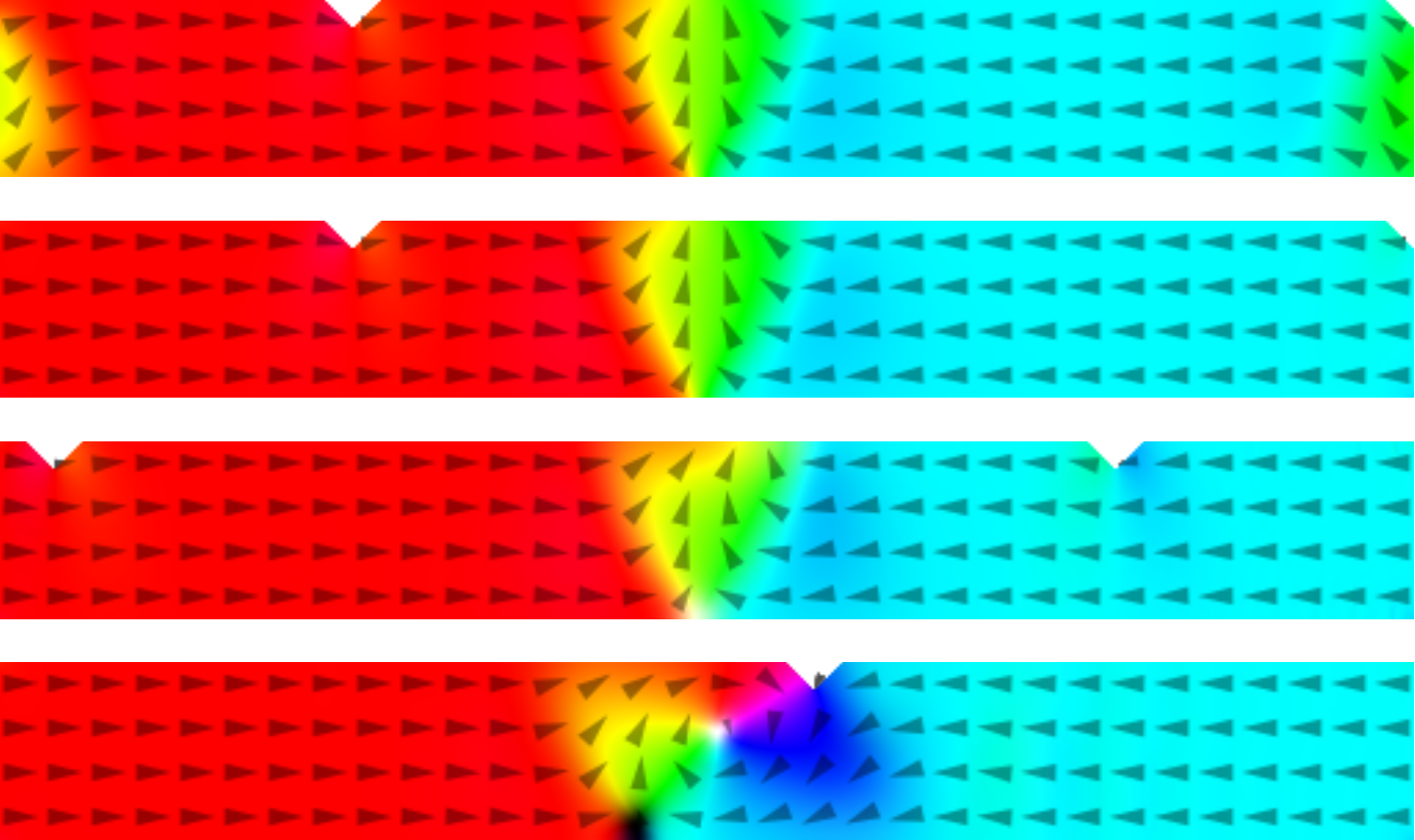}
\caption{\label{figMove} Top frame: magnetization in a 1\,$\mu$m wide, 20\,nm thick Permalloy wire of finite length. The remaining frames apply edge charge removal to simulate an infinitely long wire. The domain wall is driven by a 3\E{12}\,A/m$^2$ current while being followed by the simulation window. So that it appears steady although moving at high speed (visible by the wall breakdown). While moving, new notches enter the simulation from the right.}
\end{figure}

When following a domain wall in a long in-plane magnetized wire, we also provide the possibility to remove the magnetic charges on the ends of the wire. This simulates an effectively infinitely long wire without closure domains, as illustrated in Fig.\ref{figMove}.\\% Note that when moving the magnetization some small unequilibrium at the edges is practically unavoidable so small amounts of energy may inevitably enter or leave the system in this way.\\

Finally, when shifting the magnetization there is an option to also shift the material regions and geometry along. The geometry and material parameters for the "new" cells that enter the simulation from the side are automatically re-calculated so that new grains and geometrical features may seamlessly enter the simulation. This can be useful for, e.g., simulating a long racetrack with notches like illustrated in Fig.\ref{figMove}, or a moving domain wall in a grainy material as published in \cite{Leliaert2014}.\\

\subsection{Voronoi Tessellation}

\begin{figure}
\includegraphics[width=0.5\linewidth]{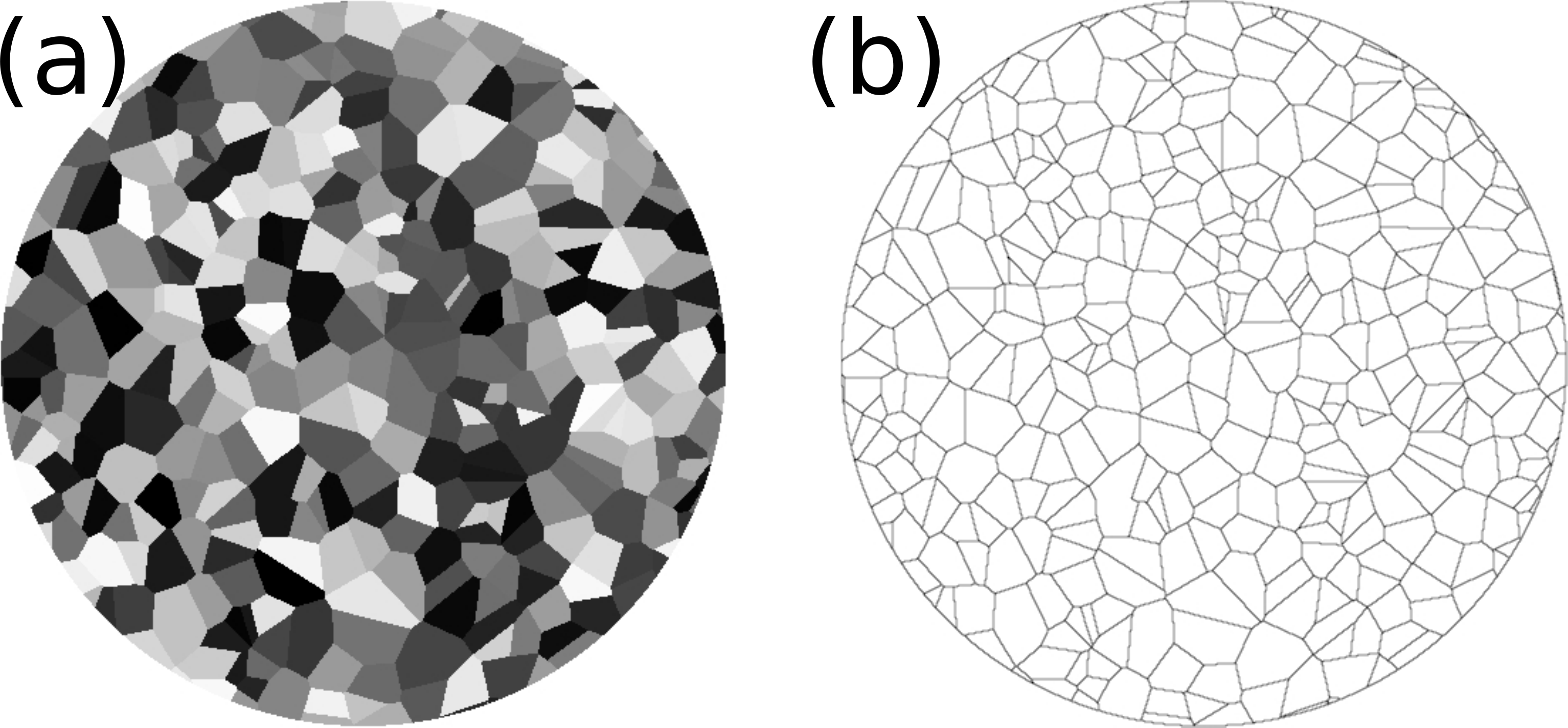}
%\tiny{\textbf{(a)}}\includegraphics[width=0.45\linewidth]{voronoi.out/regions000000}
%\tiny{\textbf{(b)}}\includegraphics[width=0.45\linewidth]{voronoi.out/ExchCoupling000000}
\caption{\label{figVoronoi}Example of a Voronoi tessellation with average 100\,nm grains in a 2048\,$\mu$m wide disk. Left: cells colored by their region index (0--256). Right: boundaries between the grains visualized by reducing the exchange coupling between them (Eq.\ref{eqBexch}), and outputting \mumax's \code{ExchCoupling} quantity, the average \Msat/\Aex\ around each cell.}
\end{figure}

\mumax provides 2D Voronoi Tessalation as a way to simulate grains in thin films, similar to OOMMF\cite{Lau2009}. It is possible to have \mumax set-up the regions map with grain-shaped islands, randomly colored with up to 256 region numbers (Fig.\ref{figVoronoi}(a)). The material parameters in each of those regions can then be varied to simulate, e.g., grains with randomly distributed anisotropy axes or even change the exchange coupling between them (Fig.\ref{figVoronoi}(b)).\\

Our implementation is compatible with the possibility to move the simulation window. E.g., when the simulation window is following a moving domain wall, new grains will automatically enter the simulation from the sides. The new grains are generated using hashes of the cell coordinates so that there is no need to store a (potentially very large) map of all the grains beyond the current simulation grid. More details can be found in\cite{Leliaert2014}\\

\subsection{Magnetic force microscopy}

\begin{figure}
\includegraphics[width=0.5\linewidth]{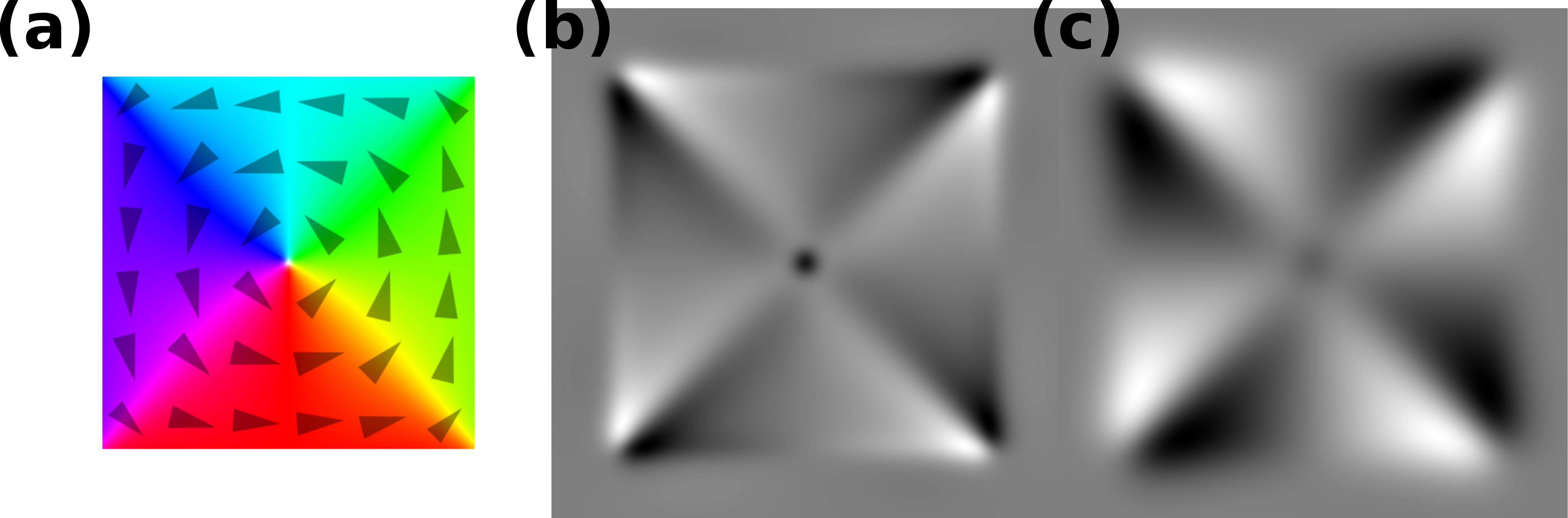}
%\tiny{\textbf{(a)}}\includegraphics[width=0.26\linewidth]{mfm.out/m000000}
%\tiny{\textbf{(b)}}\includegraphics[width=0.26\linewidth]{mfm.out/MFM000004}
%\tiny{\textbf{(c)}}\includegraphics[width=0.26\linewidth]{mfm.out/MFM000009}
\caption{\label{figMFM} (a) vortex magnetization in a 750\,nm\,\x\,750\,nm\x10\,nm Permalloy square. (b), (c) are \mumax-generated MFM images at 50\,nm and 100\,nm lift height respectively, both using AC mode and a monopole tip model.}
\end{figure}

\mumax has a built-in capability to generate magnetic force microscopy (MFM) images in Dynamic (AC) mode from a 2D magnetization. We calculate the derivative of the force between tip and sample from the convolution:

\begin{equation}
\frac{\partial F_z}{\partial z} = \sum_{i=x,y,z} M_i(x,y) * \frac{\partial^2 B_{\mathrm{tip},i}(x,y)}{\partial{z}^2} \label{eqMFM}
\end{equation}

where $\vc B_\mathrm{tip}$ is the tip's stray field evaluated in the sample plane. \mumax provides the field of an idealized dipole or monopole tip with arbitrary elevation. No attempt is made to reproduce tip fields in absolute terms as our only goal is to produce output proportional to the actual MFM contrast, like shown in Fig.\ref{figMFM}.\\

Eq. \ref{eqMFM} is implemented using FFT-acceleration similar to the magnetostatic field, and is evaluated on the GPU. Hence MFM image generation is very fast and generally takes only a few milliseconds. This makes it possible to watch "real-time" MFM images in the web interface while the magnetization is evolving in time.\\

\section{Performance}\label{perf}

\subsection{Simulation size}

Nowadays, GPU's offer massive computational performance of several TFlop/s per device. However, that compute power is only fully utilized in case of sufficient parallelization, i.e., for sufficiently large simulations. This is clearly illustrated by considering how many cells can be processed per second. I.e., $N_\mathrm{cells}/t_\mathrm{update}$ with $t_\mathrm{update}$ the time needed to calculate the torque for $N_\mathrm{cells}$ cells.  We refer to this quantity as the throughput. Given the overall complexity of $\mathcal{O}(N\log(N))$ one would expect a nearly constant throughput that slowly degrades at high $N$. For all presented throughputs, magnetostatic and exchange interactions were enabled and no output was saved.\\

\begin{figure}
\includegraphics[width=0.5\linewidth]{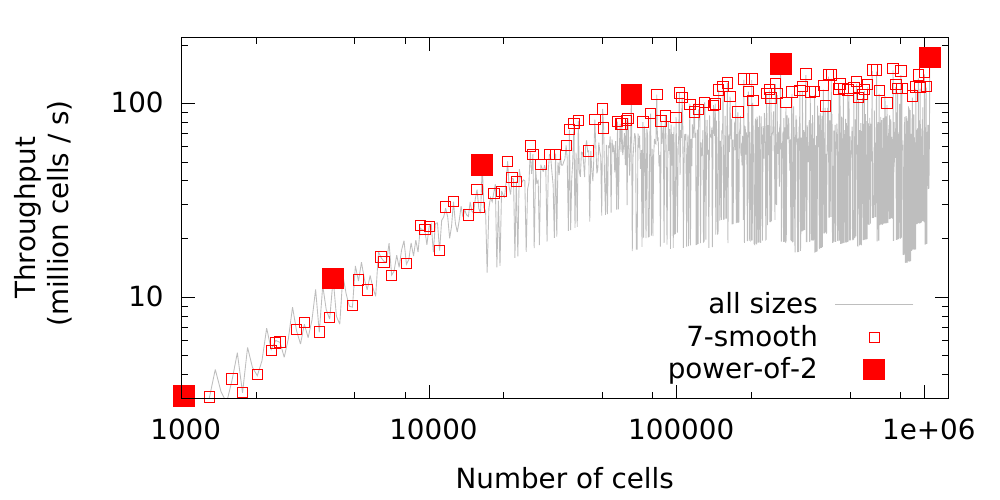}
\caption{\label{figAllsize} \mumax throughput on GTX TITAN GPU, for all $N\times N$ grid sizes up to 1024\x 1024. Numbers with only factors 2,3,5,7 are marked with an open box, pure powers of two (corresponding to Fig.\ref{figPerf}) with a full box. Proper grid sizes should be chosen to ensure optimal performance.}
\end{figure}

The throughput presented in Fig.\ref{figAllsize} for a square 2D simulation on a GTX TITAN GPU only exhibits the theoretical, nearly constant, behaviour starting from about 256\,000 cells. Below, the GPU is not fully utilized and performance drops. Fortunately, large simulations are exactly where GPU speed-up is needed most and where performance is optimal.\\

\mumax's performance is dominated by FFT calculations using the cuFFT library, which performs best for power-of-two sizes and acceptably for 7-smooth numbers (having only factors 2,3,5 and 7). Other numbers, especially primes should be avoided. This is clearly illustrated in Fig.\ref{figAllsize} where other than the recommended sizes show a performance penalty of up to about an order of magnitude. So somewhat oversizing the grid up to a nice smooth number may be beneficial to the performance.\\

Note that the data in Fig.\ref{figPerf} is for a 2D simulation. Typically a 3D simulation with the same total number of cells costs an additional factor $\propto 1.5\times$ in compute time and memory due to additional FFTs along the $z$-axis.\\

On the other hand, simulations with periodic boundary conditions will run considerably faster than their non-periodic counterparts. This is due to the absence of zero-padding which reduces FFT sizes by 2 in each periodic direction. Memory consumption will be considerably lower in this case as well.\\

\subsection{Hardware}

\begin{figure}
\includegraphics[width=0.5\linewidth]{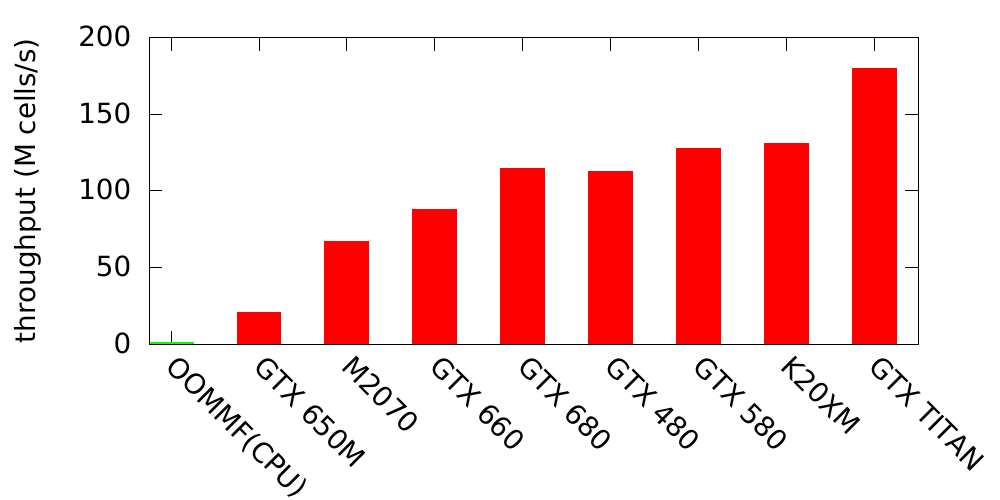}
\caption{\label{figPerf} \mumax throughput, measured in how many cells can have their torque evaluated per second (higher is better), for a 4\E{6} cell simulation (indicative for all sufficiently large simulations). For comparision, OOMMF performance on a quad-core 2.1\,GHz CPU lies around 4\,M cells/s.}
\end{figure}

Apart form the simulation size, \mumax's performance is strongly affected by the particular GPU hardware. We highlight the differences between several GPU models by comparing their throughput in Fig.\ref{figPerf}. This was done for a 4\,M cells simulation where all tested GPUs were fully utilized. So the numbers are indicative for all sufficiently large simulation sizes.\\

We also included OOMMF's throughput on a quad-core 2.1\,GHz core i7 CPU to give a rough impression of the GPU speed-up. The measured OOMMF performance (not clearly distinguishable in Fig.\ref{figPerf}) was around 4\E{6} cells/s. So with a proper GPU and sufficiently large grid sizes, a speed-up of 20--45\x with respect to a quad-core can be reached or, equivalently, a 80--180\x speed-up compared to a single-core CPU. This is in line with earlier \textsc{MuMax}1 and MicroMagnum benchmarks \cite{mumax, micromagnum}. It must be noted however that OOMMF operates in double-precision in contrast to \mumax's single-precision arithmetic, and also does not suffer reduced throughput for simulations. \\

Finally, MicroMagnum's throughput (not presented) was found to be nearly indistinguishable from \mumax. This is unsurprising since both \mumax's MicroMagnum's performance are dominated by CUDA's FFT routines. In our benchmarks on a GTX650M, differences between both packages were comparable to the noise on the timings.\\

\subsection{Memory use}

\begin{figure}
\includegraphics[width=0.5\linewidth]{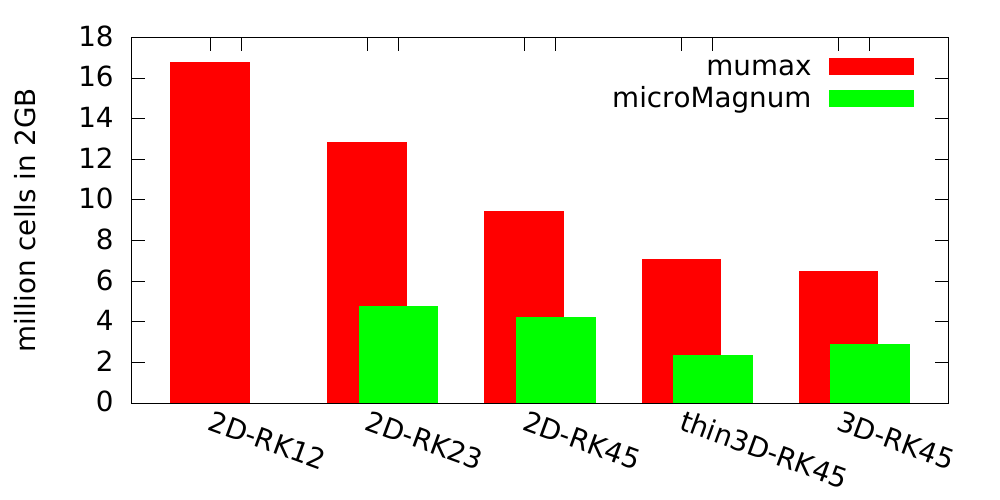}
\caption{\label{figMem} Indication of the number of cells that can be addressed with 2\,GB of GPU memory for simulations in 2D, 3D and thin 3D (here 3 layers) and using different solvers. RK45 is \mumax's default solver for dynamics, RK23 for relaxation. Only magnetostatic, exchange and Zeeman terms were enabled.}
\end{figure}

In contrast to their massive computational power, GPUs are typically limited to rather small amounts of memory (currently 1---6\,GB). Therefore, \mumax was heavily optimized to use as little memory as possible. E.g., we exploited the magnetostatic kernel symmetries and zero elements and make heavy use of memory pooling and recycling.\\

Also, \mumax employs minimal zero-padding in the common situation of 3D simulations with only a small number of layers. For up to 10 layers there is no need to use a power of two, and memory usage will be somewhat reduced as well.\\

In this way, \mumax on a GPU with only 2\,GB of memory is able to simulate about 9 million cells in 2D and 6 million in 3D, or about 2\,\x more than MicroMagnum v0.2\cite{micromagnum} (see Fig.\ref{figMem}). When using a lower-order solver this number can be further increased to 12\E6 cells with RK23 (2D) or 16\E6 cells with RK12(2D), all in 2\,GB. Cards like the GTX TITAN and K20XM, with 6\,GB RAM can store proportionally more, e.g., 31\,M cells for 2D with the RK45 solver.\\

\section{Conclusion}

We have presented in detail the micromagnetic model employed by \mumax, as well as a verification for each of its components. GPU acceleration provides a speed-up of 1--2 orders of magnitude compared to CPU-based micromagnetic simulations. In addition, \mumax's low memory requirements open up the possibility of very large-scale micromagnetic simulations, a regime where the GPU's potential is fully utilized and where the speed-up is also needed most. E.g., depending on the solver type \mumax can fit 10--16 million cells in 2\,GB GPU RAM --- about 2\x more than MuMax2 or MicroMagnum.\\

\mumax is open-source and designed to be easily extensible, so anybody can in principle add functionality. Some extensions like a moving simulation window, edge charge removal, Voronoi tessellation and MFM images have been permanently merged into \mumax and more extensions are expected in the future.

\section{Acknowledgements}

This work was supported by the Flanders Research Foundation (FWO).\\

The authors would like to cordially thank Ahmad Syukri bin Abdollah, Alex Mellnik, Aurelio Hierro, Ben Van de Wiele, Colin Jermain, Damien Louis, Ezio Iacocca, Gabriel Chaves, Graham Rowlands, Henning Ulrichs, Joo-Von Kim, Lasse Laurson, Mathias Helsen, Raffaele Pellicelli, Rémy Lassalle-Balier, Robert Stamps and Xuanyao (Kelvin) Fong for the fruitful discussions, contributions or feedback, as well as all others who tested early \textsc{MuMax} versions.\\

\mumax uses svgo (\url{http://github.com/ajstarks/svgo}), copyright Anthony Starks, and freetype-go (\url{http://code.google.com/p/freetype-go}), copyright Google Inc., Jeff R. Allen, Rémy Oudompheng, Roger Peppe.

\bibliography{bibliography}

\clearpage

\appendix

\section{Input scripts}\label{appendixA}

\subsection{Geometry (\Fig{figCSG})}
\begin{verbatim}
Nx := 256; Ny := 128; Nz := 32
c  := 4e-9
setgridsize(Nx, Ny, Nz)
setcellsize(c, c, c)

s := ellipsoid(c*Nx/2, c*Ny/2, c*Nz).transl(-Nx/7*c, 0, 0)
s  = s.add(rect(c*Nx/3, c*Ny/2).rotz(20*pi/180).transl(c*Nx/6, 0, 0))
setgeom(s)

msat = 800e3
aex  = 13e-12
m    = uniform(1, 0, 0)

relax()
save(m)
\end{verbatim}

\subsection{Precession (\Fig{figtqLL})}
\begin{verbatim}
SetGridSize(1, 1, 1)
SetCellSize(1e-9, 1e-9, 1e-9)
m = Uniform(1, 0, 0)
B_ext = Vector(0, 0, 0.1)
EnableDemag = false
TableAutosave(10e-12)
Run(3e-9)
\end{verbatim}

\subsection{Cube demag tensor (Table\,\ref{tabCube})}
\begin{verbatim}
Msat = 1 / mu0          // 1 Tesla

// prolate cells
for p:=0; p < 7; p++{
    aspect := pow(2, p)

    setGridSize(aspect, aspect, 1)
    setCellSize(1/aspect, 1/aspect, 1)

    m = uniform(1, 0, 0)
    Kx := B_demag.Average()

    m = uniform(0, 1, 0)
    Ky := B_demag.Average()

    m = uniform(0, 0, 1)
    Kz := B_demag.Average()

    fprintln("result.txt", "1/", aspect, Kx.x(), Ky.y(), Kz.z())
}

// elongated ceels
for p:=0; p < 4; p++{
    aspect := pow(2, p)

    setGridSize(1, 1, aspect)
    setCellSize(1, 1, 1/aspect)

    m = uniform(1, 0, 0)
    Kx := B_demag.Average()

    m = uniform(0, 1, 0)
    Ky := B_demag.Average()

    m = uniform(0, 0, 1)
    Kz := B_demag.Average()

    fprintln("result.txt", aspect, "/1", Kx.x(), Ky.y(), Kz.z())
}


\end{verbatim}

\subsection{Long-range demag (\Fig{figLong})}
\begin{verbatim}
setGridSize(512, 1, 1)
setCellSize(1e-9, 1e-9, 1e-9)

SetGeom(cell(0, 0, 0))
Msat = 1/mu0

m = uniform(1, 0, 0)
save(B_demag)
\end{verbatim}

\subsection{Sheet demag tensor with PBC (Table\,\ref{tabPBC1})}
\begin{verbatim}
c := 1e-9
SetCellSize(c, c, c)

Msat = 1/mu0
m    = uniform(0, 0, 1)
N   := 2

// small grid with PBC
for j := 0; j<10; j++{
    i := pow(2, j)
    SetGridSize(N, N, 1)
    SetPBC(i, i, 0)
    fprintln("pbc.txt", i, B_demag.Average())
}


// equivalent large grid without PBC
for j := 0; j<10; j++{
    i := pow(2, j)
    SetGridSize(i*2*N, i*2*N, 1)
    SetPBC(0, 0, 0)
    // evaluate over central slab
    defRegion(0, universe())
    defRegion(1, rect(c*N, c*N))  
    fprintln("nopbc.txt", i, B_demag.Region(1).Average())
}
\end{verbatim}

\subsection{Rod demag tensor with PBC (Table\,\ref{tabPBC2})}
\begin{verbatim}
c := 1e-9
SetCellSize(c, c, c)

Msat = 1/mu0
m    = uniform(0, 0, 1)
N   := 2

// small grid with PBC
for j := 0; j<10; j++{
    i := pow(2, j)
    SetGridSize(N, N, 1)
    SetPBC(0, 0, i)
    fprintln("pbc.txt", i, B_demag.Average())
}

// equivalent large grid without PBC
for j := 0; j<10; j++{
    i := pow(2, j)
    SetGridSize(N, N, 1*i*2)
    SetPBC(0, 0, 0)
    
    // evaluate over central part
    defRegion(0, universe())
    defRegion(1, cuboid(c*N, c*N, c*N))
    fprintln("nopbc.txt", i, B_demag.Region(1).Average())
}
\end{verbatim}

\subsection{Exchange energy (\Fig{figExchE})}
\begin{verbatim}
N := 1024
c := 1e-9
SetGridSize(N, 1, 1)
SetCellSize(c, c, c)
SetPBC(1, 0, 0)
vol := N * c * c * c

A  := 10e-12
Aex = A
Msat = 1e6
EnableDemag = false

mag := newSlice(3, N, 1, 1)
for k:=0; k<N/2; k++{
    for i:=0; i<N; i++{
        phase := (i * k * 2*pi) / N
        my := sin(phase)
        mz := cos(phase)
        mag.set(1, i, 0, 0, my)
        mag.set(2, i, 0, 0, mz)
    }
    m.SetArray(mag)
    save(m)
    
    Energy := vol * A * pow( 2*pi*k / (N*c), 2 ) 
    fprintln("output.txt", k, Energy, E_exch)
}
\end{verbatim}

\subsection{DM interaction (\Fig{figThiaville})}
\begin{verbatim}
/*
    Test for the DMI interaction, inspired by Thiaville et al. EPL 100 2012.
    In this simulation, the DMI interaction converts a Bloch wall to a Néel wall.
*/

SetGridSize(128, 128, 1)
SetCellSize(250e-9/128, 250e-9/128, 0.6e-9)

Msat  = 1100e3
Aex   = 16e-12
alpha = 3
AnisU = vector(0, 0, 1)
Ku1   = 1.27E6
m     = TwoDomain(0, 0, -1, 1, 1, 0, 0, 0, 1) // down-up domains with wall between Bloch and Néél type

tableadd(Dex)
for d:=0.0; d<0.3e-3; d+=0.001e-3{
    Dex = d
    relax()
    tablesave()
}
\end{verbatim}

\subsection{Uniaxial anisotropy (\Fig{figAnis})}
\begin{verbatim}
setGridSize(1, 1, 1)
setCellSize(1e-9, 1e-9, 1e-9)
V := pow(1e-9, 3)

Msat  = 1000e3
AnisU = vector(1, 0, 0)

// vary magnetization direction
for i:=0; i<=360; i++{
    theta := i*pi/180
    m = uniform(cos(theta), sin(theta), 0)

    Ku1 = 1e6; Ku2 = 0
    fprintln("easy.txt", theta, E_anis.get()/V) 	

    Ku1 = -1e6; Ku2 = 0
    fprintln("hard.txt", theta, E_anis.get()/V) 	

    Ku1 = 0; Ku2 = 1e6
    fprintln("easy2.txt", theta, E_anis.get()/V) 	

    Ku1 = 0; Ku2 = -1e6
    fprintln("hard2.txt", theta, E_anis.get()/V) 	
}
\end{verbatim}

\subsection{Cubic anisotropy (\Fig{figAnis})}
\begin{verbatim}
setGridSize(1, 1, 1)
setCellSize(1e-9, 1e-9, 1e-9)
V := pow(1e-9, 3)

Msat   = 1000e3
AnisC1 = vector(1, 0, 0)
AnisC2 = vector(0, 1, 0)

// vary magnetization direction
for i:=0; i<=360; i++{
    theta := i*pi/180
    m = uniform(cos(theta), sin(theta), 0)

    Kc1 = 1e6; Kc2 = 0; Kc3 = 0
    fprintln("easy.txt", theta, E_anis.get()/V)	

    Kc1 = -1e6; Kc2 = 0; Kc3 = 0
    fprintln("hard.txt", theta, E_anis.get()/V)	

    Kc1 = 0; Kc2 = 1e6; Kc3 = 0
    fprintln("easy2.txt", theta, E_anis.get()/V)	

    Kc1 = 0; Kc2 = -1e6; Kc3 = 0
    fprintln("hard2.txt", theta, E_anis.get()/V)	

    Kc1 = 0; Kc2 = 0; Kc3 = 1e6
    fprintln("easy3.txt", theta, E_anis.get()/V)	

    Kc1 = 0; Kc2 = 0; Kc3 = -1e6
    fprintln("hard3.txt", theta, E_anis.get()/V)	
}

\end{verbatim}

\subsection{Thermal fluctuations (\Fig{figSwitch})}
\begin{verbatim}
c := 10e-9
setcellsize(c, c, c)
setgridsize(512, 512, 1)

Msat  = 1e6
Aex   = 0
alpha = 0.1
AnisU = vector(0, 0, 1)
m     = uniform(0, 0, 1)
fixdt = 1e-12
Temp  = 100              // varied in batch
Ku1   = 1e4              // varied in batch
enabledemag = false
setsolver(2)

run(10e-9)               // reach equilibrium
autosave(m, .1e-6)       // will count number of flips in post-processing
run(1e-6)
\end{verbatim}

\subsection{Slonzewski STT (\Fig{figStd5b})}
\begin{verbatim}
length := 160e-9 ;      Nx := 64
width  :=  80e-9 ;      Ny := 32
thick  :=  5e-9  ;      Nz := 1

Msat    = 800e3  ;      Pol          = 0.5669
Aex     = 13e-12 ;      Lambda       = 2
alpha   = 0.01   ;      EpsilonPrime = 1

SetGridSize(Nx, Ny, Nz)
SetCellSize(length/Nx, width/Ny, thick/Nz)

current:= -0.006 // Ampere
J       =  vector(0, 0, current/(length*width))

theta     := pi*20/180   // Direction of fixed layer
FixedLayer = vector(cos(theta), sin(theta), 0)
m          = uniform(1, 0, 0)

tableautosave(10e-12)
run(2e-9)
\end{verbatim}

\subsection{Solver convergence and MaxErr (Figs.\,\ref{figConvergence} and \ref{figMaxErr})}
\begin{verbatim}
SetGridSize(1, 1, 1)
SetCellSize(1e-9, 1e-9, 1e-9)

B          := 0.1 // Tesla
B_ext       = vector(0, 0, B)
Msat        = 1/mu0
Aex         = 10e-12
EnableDemag = false

SetSolver(5)  //varied

for i:=1e-8; i<1e-2; i=i*sqrt(2){
    maxErr = i

    N0 := Step
    t = 0
    m = Uniform(1, 0, 1)
    runtime := 2*pi/(B*1.7595e11) // 1 period
    run(runtime)

    m1 := m.average()
    m0 := vector(1/sqrt(2), 0, 1/sqrt(2)) // analytical solution
    error := m0.sub(m1).len()
    fprintln("result.txt", i, runtime/(Step-N0), error)
}

\end{verbatim}

\subsection{Standard Problem 1 (\Fig{figStd1})}
\begin{verbatim}
setGridSize(512, 256, 1)
setCellSize(2e-6/512, 1e-6/256, 20e-9)

Aex   = 1.3e-11 
Msat  = 8.0e5 
Ku1   = 5.0e2 
AnisU = vector(1, 0.001, 0) // easy axis nominally parallel to the long edge 

// hysteresis sweep
Bmax := 75e-3
tableadd(B_ext)

// virgin curve
m = vortex(1, 1).add(0.9, randommag())
for B := 0.0; B <= Bmax; B+=1e-3{
    B_ext = vector(B, 0, 0)
    relax()
    tableSave()
}

// sweep down
for B := Bmax; B >= -Bmax; B-=1e-3{
    B_ext = vector(B, 0, 0)
    relax()
    tableSave()
}

// sweep up
for B := -Bmax; B <= Bmax; B+=1e-3{
    B_ext = vector(B, 0, 0)
    relax()
    tableSave()
}
\end{verbatim}

\subsection{Standard Problem 2 (Figs.\,\ref{figStd2rem} and \ref{figStd2hc})}
\begin{verbatim}
// Msat and Aex should not matter
Msat = 1000e3           
Aex  = 10e-12

// exchange length
lex := sqrt(Aex.GetRegion(0) / (0.5 * mu0 * pow(Msat.GetRegion(0),2)))

mx := m.comp(0)
my := m.comp(1)
mz := m.comp(2)

// scan magnet size d
for d := 30; d >= 1; d--{
    Sizex := 5*lex*d
    Sizey := 1*lex*d
    Sizez := 0.1*lex*d

    // choose cells no larger than 0.5*lex
    nx := pow(2, ilogb(Sizex / (5*0.5*lex))) * 5
    ny := nx / 5
    SetGridSize(nx, ny, 1)
    SetCellSize(Sizex/nx, Sizey/ny, Sizez)

    // find remanence
    m = uniform(1, 0.3, 0)
    relax()
    remanence := m.average()

    // find coercivity by scanning applied field
    bc := 0.0
    Ms := Msat.GetRegion(0)
    for bc = 0.0445*Ms; (mx.average()+my.average()+mz.average()) > 0; bc += 0.00005*Ms{
        B_ext = vector(-bc*mu0/sqrt(3), -bc*mu0/sqrt(3), -bc*mu0/sqrt(3))
        relax()
    }
    B_ext = vector(0, 0, 0)

    fprintln("table.txt", d, remanence.x(), remanence.y(), bc/Ms)
}

\end{verbatim}

\subsection{Standard Problem 3 (\Fig{figStd3})}
\begin{verbatim}
N := 16
setGridSize(N, N, N)
setCellSize(1e-9, 1e-9, 1e-9) // will be varied

Msat  = 500e3   // Msat and Aex should not matter
Aex   = 20e-12

Ms   := Msat.GetRegion(0)
A    := Aex.GetRegion(0)
Km   := 0.5 * mu0 * Ms * Ms
lex  := sqrt(A / Km)
AnisU = Vector(0, 0, 1)
Ku1   = 0.1 * Km

// scan edge length L
for L := 8.0; L<=9.0; L+=0.02{
    c := L*lex/N
    setCellSize(c, c, c)
    m.loadFile("std3-flower.ovf") // load initial state
    V := pow(c*N, 3)
    relax()
    m_:= m.Average()
    fprintln("result.txt", L, m_.X(), m_.Y(), m_.Z(), E_total.Get(), E_exch.Get(), E_demag.Get(), E_anis.Get(), Km, V)
}

\end{verbatim}

\subsection{Standard Problem 4 (\Fig{figStd4})}
\begin{verbatim}
Nx:=200
Ny:=50
setgridsize(Nx, Ny, 1)
setcellsize(500e-9/Nx, 125e-9/Ny, 3e-9)

Msat  = 800e3
Aex   = 13e-12
alpha = 0.02

m = uniform(1, .1, 0)
relax()

tableautosave(5e-12)

B_ext = vector(-24.6E-3, 4.3E-3, 0)
mx := m.comp(0)
runWhile(mx.average() > 0)
save(m)
run(1e-9)

\end{verbatim}

\subsection{Standard Problem 5 (\Fig{figStd5})}
\begin{verbatim}
Nx := 50
Ny := 50
Nz := 5
setgridsize(Nx, Ny, Nz)
setcellsize(100e-9/Nx, 100e-9/Nx, 10e-9/Nz)

Msat  = 800e3
Aex   = 13e-12
alpha = 0.1
xi    = 0.05

m = vortex(1, 1)
relax()
saveas(m, "relaxed.ovf")

J     = vector(1e12, 0, 0)
Pol   = 1

tableAutoSave(5e-12)
run(10e-9)
\end{verbatim}

\subsection{Extension: moving reference frame (\Fig{figMove})}
\begin{verbatim}
Nx := 512
Ny := 64
c := 2e-9
setGridSize(Nx, Ny, 1)
setCellSize(c, c, 20e-9)

w := Nx * c
h := Ny * c

Msat    = 800e3
Aex     = 13e-12
Xi      = 0.1
alpha   = 0.02 
m       = twodomain(1,.1,0, 0,1,0, -1,-.1,0)

notch := rect(15*c, 15*c).rotZ(45*pi/180).transl(0, Ny*c/2, 0).repeat(Nx*c/2 + 2*Ny*c, 0, 0).inverse()
setgeom(notch.transl(-2*Ny*c, 0, 0))

snapshotformat = "PNG"

relax()
snapshot(m)

// Remove surface charges from left (mx=1) and right (mx=-1) sides to mimic infinitely long wire. We have to specify the region (0) at the boundaries.
BoundaryRegion := 0
MagLeft        := 1
MagRight       := -1
ext_rmSurfaceCharge(BoundaryRegion, MagLeft, MagRight)

relax()
snapshot(m)

ext_centerWall(0) // keep m[0] (m_x) close to zero

// Schedule output
autosnapshot(m, 50e-12)
tableadd(ext_dwpos)   // domain wall position
tableautosave(10e-12)

// Run the simulation with current through the sample
pol = 0.56
J   = vector(-3e12, 0, 0)
Run(5e-9)

\end{verbatim}

\subsection{Extension: Magnetic Force Microscopy (\Fig{figMFM})}
\begin{verbatim}
setgridsize(512, 512, 1)
setcellsize(2e-9, 2e-9, 10e-9)
setgeom(rect(750e-9, 750e-9))

msat    = 800e3
aex     = 13e-12
m       = vortex(1, 1)

relax()
save(m)

// MFM images for various lift heights
for i:=10; i<=200; i+=10{
    MFMLift = i*1e-9
    Snapshot(MFM)
}
\end{verbatim}

\subsection{Benchmark: throughput (Figs.\,\ref{figAllsize} and \ref{figPerf})}
\begin{verbatim}
msat  = 800e3
aex   = 13e-12
alpha = 0.01
b_ext = vector(0, 0.01, 0)
m     = uniform(1, 0, 0)

setcellsize(4e-9, 4e-9, 4e-9)

// vary grid sizes
for n:=16; n<=1024; n++{
    setgridsize(n, n, 1)
    N2 := n*n
    steps(1) // warm-up kernel

    t = 0
    start := now()
    neval0 := Neval()

    steps(100)

    wall := since(start).Seconds() 
    nevl := Neval() - neval0
    fprintln("benchmark.txt", N2, N2*nevl/wall, t/nevl)
}
\end{verbatim}

\subsection{Benchmark: memory (\Fig{figMem})}
\begin{verbatim}
msat  = 800e3
aex   = 13e-12
alpha = 0.01
setcellsize(4e-9, 4e-9, 4e-9)

// increase grid size until we run out of memory
for e:=5; e<100; e++{
    n := 128*e
    setgridsize(n, n, 1)
    steps(1)                 // will crash here if out of memory
    fprintln("mem.txt", n*n) // save number of cells
}
\end{verbatim}

\end{document}